\documentclass{article}
\pdfoutput=1

\usepackage{arxiv}

\usepackage[utf8]{inputenc} % allow utf-8 input
\usepackage[T1]{fontenc}    % use 8-bit T1 fonts
\usepackage{hyperref}       % hyperlinks
\usepackage{url}            % simple URL typesetting
\usepackage{booktabs}       % professional-quality tables
\usepackage{amsfonts}       % blackboard math symbols
\usepackage{nicefrac}       % compact symbols for 1/2, etc.
\usepackage{microtype}      % microtypography
\usepackage{lipsum}
\usepackage{amsmath}
\usepackage{array}
\usepackage{mathtools}
\usepackage{bm}

\usepackage{graphicx}
\graphicspath{{images/}{../images/}}

\usepackage{cite}

\usepackage{enumitem}
\setlist[enumerate]{leftmargin=.5in}
\setlist[itemize]{leftmargin=.5in}

\newtheorem{definition}{Definition}[section]
\newtheorem{definition*}{Definition}

\title{Hybrid Function Representation for Heterogeneous Objects}

\author{
  A. Tereshin \\
  National Centre for Computer Animation\\
  Bournemouth University\\
  Poole, United Kingdom\\
  \texttt{atereshin@bournemouth.ac.uk} \\
  \And
  A. Pasko \\
  Skolkovo Institute of Science and Technology\\
  Skolkovo, Russia \\
  \texttt{a.pasko@skoltech.ru} \\
  \And
  O. Fryazinov \\
  National Centre for Computer Animation \\
  Bournemouth University \\
  Poole, United Kingdom \\
  \texttt{ofryazinov@bournemouth.ac.uk}\\
  \And
  V. Adzhiev \\
  National Centre for Computer Animation \\
  Bournemouth University \\
  Poole, United Kingdom \\
  \texttt{vadzhiev@bournemouth.ac.uk}
}

\begin{document}
\maketitle

\begin{abstract}
    Heterogeneous object modelling is an emerging area where geometric shapes are considered in concert with their internal physically-based attributes. This paper describes a novel theoretical and practical framework for modelling volumetric heterogeneous objects on the basis of a novel unifying functionally-based hybrid representation called HFRep. This new representation allows for obtaining a continuous smooth distance field in Euclidean space and preserves the advantages of the conventional representations based on scalar fields of different kinds without their drawbacks. We systematically describe the mathematical and algorithmic basics of HFRep. The steps of the basic algorithm are presented in detail for both geometry and attributes. To solve some problematic issues, we have suggested several practical solutions, including a new algorithm for solving the eikonal equation on hierarchical grids. Finally, we show the practicality of the approach by modelling several representative heterogeneous objects, including those of a time-variant nature.
\end{abstract}

% keywords can be removed

\section{Introduction}
\label{sec:intro}

Heterogeneous volumetric object modelling is a rapidly developing field and has a variety of different applications. Volume modelling is concerned with computer representation of object surface geometry as well as its interior. Homogeneous volume modelling, better known as solid modelling, deals with volume interior uniformly filled by a single material. Heterogeneous object is a volumetric object with interior structure where different physically-based attributes are defined, e.g. spatial different material compositions, micro-structures, colour, density, etc. \cite{KOU:2007, Li:2020}. This type of objects is widely used in applications where the presence of the interior structures is an important part of the model. Additive manufacturing, physical simulation and visual effects are examples of such applications.

The most widely used representations for defining heterogeneous objects are boundary representation, distance-based representations, function representation and voxels. Boundary representation (BRep) \cite{Lei:2014} maintains its prevailing role due to its numerous well-known advantages. It works well in solid modelling for objects consisting of a set of polygonal surface patches stitched together to envelope the uniform and homogeneous structure of its material. However, BRep is not inherently natural for dealing with heterogeneous objects, especially in the context of additive manufacturing and 3D printing \cite{Livensu:2017}, where volume-based multi-material properties are paramount as well as in physical simulation where the exact representation rather than an approximate one can be important \cite{Nealen:2006}. 

On the contrary, volumetric representations in the form of voxels \cite{Wang:2011} are more natural for defining such heterogeneous objects as they are based on volumetric grids. Voxels represent an object as a set of cubic cells at which the geometry along with the object attributes are defined. However, this representation essentially approximates both the geometry model  and the material distribution in interior of the object as their definition is limited by the resolution of the voxel grid. 

On the other hand, function-based, and more specifically, distance-based  representations are able to represent the object and its interior structure in both continuous and discrete forms \cite{Jones:2006}. They are exact, embrace a wide range of geometric shapes and naturally define many physically-based attributes. There are a lot of well-established operations for these representations. Most of them provide distances to the object surface. However, distance functions (DFs) are not essentially continuous, they can have medial gradient discontinuities and are not necessarily smooth. This potentially results in non-watertight surfaces, and in artefacts, such as creases, after applying some operations, for instance, blending and metamorphosis, which are important for many applications. Undesired artefacts (stresses, creases, etc.) can also appear as the result of defining  distance-based attribute functions.

We consider function-based and distance-based representations as a promising conceptual and practical scheme to deal with heterogeneous objects, especially in the context of a number of topical application areas concerned with exact volume-based geometric modelling, animation, simulation and fabrication. However, the existing representational schemes of that type appear in many variations and the field as a whole exhibits a rather fragmented and not properly formalised suite of methods. There is an obvious need for a properly substantiated and unifying theoretical and practical framework. This challenge can be considered in the context of the emergence of new representational paradigms suitable for the maturing applications, such as modelling of material structures, that was outlined and substantiated in \cite{Regli:2016}.

In this work we propose a novel function-based representational scheme. We introduce a mathematical framework called hybrid function representation for defining a heterogeneous volumetric object with its attributes in continuous and discrete forms. It is based on hybridisation of several DF-based representations that unifies their advantages and compensates for their drawbacks. This representational scheme aims at dealing with heterogeneous objects with some specific time-variant properties important in physical simulations related to both geometry and attributes. The idea was initially tested in a short paper \cite{Tereshin:2019} where the scheme unifying the function representation (FRep) and the signed distance functions (SDFs) had been sketchily outlined.

The contributions of our work can be formulated as follows:
\begin{itemize}
    \item We provide a thorough survey of the relevant representations aiming at their classification and identifying their advantages and drawbacks. We formalise the notions of the adaptively sampled distance fields (ADFs) and interior distance fields (IDFs).
    
    \item On the basis of an analysis of the well-established FRep and DF-based representations, namely SDFs, ADFs and IDFs, we formulate the requirements for a novel unifying hybrid representation called HFRep.
    
    \item We propose a mathematically substantiated theoretical description of the HFRep with an emphasis on defining functions for HFRep objects' geometry and attributes.
    
    \item We describe a basic algorithm for generating HFRep objects in terms of their geometry and attributes, and develop its main steps in a detailed step-by-step manner.
    
    \item We identify the problematic issues associated with several steps of the basic algorithm and propose several practical solutions. In particular, we present a novel hierarchical fast iterative method for solving the eikonal equation on hierarchical grids in 2D. The developed algorithm was used for generating HFRep based on FRep and ADF.
\end{itemize}

\section{Related works}
\label{sec:related_w}

There is a huge body of works dealing with different aspects of representational schemes for volumetric heterogeneous objects. In this section we first concentrate on those works that deal with representations for geometric shapes. Then we consider some existing hybrid representations. The basic methods for defining attributes in interior of volumetric objects are also reviewed.

\subsection{Geometry representations}
\label{ssec:geom_reps}

An overall object geometry can be represented by boundary surfaces or by any other solid representational scheme including procedurally defined scalar fields.

Boundary representation (BRep) remains the most popular representation. It can be described by a polygonal or other surface model. Polygonal models can be represented as nested polygonal meshes bounding the regions with different material density values \cite{Lei:2014}. This representation scheme has the following problems: the possible presence of holes or gaps in a mesh, normals can be flipped, triangles in the mesh can be intersecting or overlapping with each other, polygonal shells can be noisy. Among many polygon-based approaches applicable to heterogeneous objects we pay attention to the diffusion surfaces introduced in \cite{Takayama:2010}. This approach deals with 3D surfaces with colours defined on both sides, such that the interior colours in the volume are obtained by diffusing colors from nearby surfaces. It was used for modelling objects with rotational symmetry. It is efficient to compute, but cross-sections of the mesh obtained with further triangulation could suffer from discretisation artefacts. 

Another way to represent the overall object geometry is constructive solid geometry (CSG) \cite{Requicha:1980}. Originally all solids were homogeneous, but later primitives could carry on some information that can be interpreted as a material index \cite{Bowyer:1995}. The operations on attributes corresponding to set-theoretic operations were provided.

The most widely used method for defining heterogeneous objects is the voxel representation \cite{Wang:2011, Bader:2018}. The object is subdivided into multiple cubic cells with defined geometric and attribute parts in them. However, geometric and attribute properties are essentially approximated according to the voxel grid resolution.

In the context of this work we pay a special attention to defining a heterogeneous object geometry using different types of scalar fields. The most common schemes of that type are already mentioned FRep \cite{Pasko:1995}, SDFs \cite{Jones:2006}, ADFs \cite{Frisken:2000} as well as the shape aware distance fields which are represented by functions that we call interior distance functions (IDFs). We will discuss them in more details in the next section.

Another widely used approach for obtaining a continuous distance based
definition of the object is to compute the solution of the optimal mass transportation \cite{Solomon:2014}. This method assumes the numerical solution of a partial differential equation (PDE) dedicated to the Monge-Kantorovich optimisation problem which can be quite time-consuming. In \cite{Zhan:2012}, volumetric objects with multiple internal regions were suggested to define the object-space multiphase implicit functions. These functions preserve sharp features of the object and in some cases provide better results than SDFs.

Distance-based objects can be also defined using the level-set method \cite{Gibou:2018} which provides an implicit representation of a moving front. The main advantage of this method is that it could handle various topological changes of the object thus implementing the dynamic implicit surfaces. The evolution of the front is controlled by the solution of the level-set equation. The obtained function is transformed into a signed distance function using the solution of some reinitialisation equation. Level-set methods have been used in many applications, such as shape optimisation, computational fluid dynamics, trajectory planning, image processing and others \cite{Gibou:2018}.

\subsection{Hybrid representations}
\label{ssec:hybryd_reps}

The main feature of any hybrid representation is that it unifies advantages of several representations and compensate for their drawbacks. In \cite{Pasko:2001}, the concept of hypervolumes was introduced as an extension of the general object model \cite{KUMAR:1999} that unifies the advantages of FRep and hybrid volumes. Hypervolume describes a heterogeneous object as n-dimensional point-set with defined attributes, operations and relations over them. Another hybrid approach called hybrid surface representation was introduced in \cite{Kim:2004}. It is based on BRep and an implicit surface representation (V-Rep) and was used for heterogeneous volumetric modelling and sculpting. 

There are approaches when an entire object can be split into disjoint or adjacent components sharing their boundaries. The space partitions can be defined by additional boundary surfaces or scalar fields. In the most general case, these partitions are represented by mixed-dimensional cells combined into a cell complex. The combination of a cellular representation and a functionally based constructive representation was proposed in \cite{Adzhiev:2002}. This model makes it possible to represent dimensionally non-homogeneous elements and their cellular representations. The authors showed that attributes may reflect not only material, but any volumetric distribution such as density or temperature.

There are some works dedicated to the construction of hybrid representations based on SDFs, ADFs and IDFs. In \cite{Tsukanov:2011}, the authors have introduced hybridisation of meshfree, RBF-based, DF-based and collocating techniques for solving engineering analysis problems. The proposed technique enables exact treatment of all boundary conditions and can be used with both structured and unstructured grids. In patent \cite{Sullivan:2015}, Sullivan has introduced the hybrid ADFs which represented the object by a set of cells. In work \cite{Allegre:2004}, the authors introduced a new structure called HybridTree that is based on an extended CSG tree which unifies advantages of skeletal implicit surfaces and polygonal meshes. The hybrid biharmonic distances that are defined similarly to diffusion and commute-time (graph) distances were introduced in \cite{Lipman:2010} for solving some shape analysis tasks.  In \cite{Kim:2015}, a concept of the hybrid ADF was introduced for the detailed representation of the dynamically changing liquid-solid mixed surfaces. 

\subsection{Material and attribute representations}
\label{ssec:mat_attr_reps}

A notable early framework called constructive volume geometry (CVG) for modelling heterogeneous objects using scalar fields was decsribed in \cite{Chen:2000}. The CVG algebraic representation describes both interior and exterior of the object that using regular or hierarchical data-structures. The CVG mathematical framework works with spatial objects defined as a tuple $O=(F_O, A_1, ..., A_n)$, where $F_O$ is an opacity field that $F_O: \mathbb{R}^3 \mapsto [0,1]$ and $A_i$ are attribute fields. The opacity field defined by the function $F_O$ is non-distance based and it is not essentially continuous. Discrete fields also can be used in this representation using some interpolation procedure.

Multi-material heterogeneous volumetric objects \cite{Tuan:2018} consist of three elements: object geometry, object components (e.g. domains, partitions or cells sharing their boundaries) and material distribution. Material distributions can be defined using material indexes, piecewise polynomials or continuous scalar fields that provide a resolution independent distribution.

To define material in the interior of the object, a spatial partitioning of the object in several spatial regions should be made. Perhaps, the most widely used approach is a voxel representation of the object \cite{Wang:2011}. In \cite{Hiller:2009}, Hiller and Lipson suggested to use a voxel data structure as a material building-block for layered manufacturing. In another work \cite{Doubrovski:2015}, a bitmap voxel-based method that uses multi-material high-resolution additive manufacturing (AM)  was introduced. The material properties are combined in local material compositions that are further fetched in a AM system. In \cite{Bader:2018} a multi-material voxel-printing method using a high-resolution dithering technique was introduced. The material in the voxelised object is defined using spatial indexing. 

Material distribution in interior of the volumetric object can also be defined using DF-based approaches. In \cite{BISWAS:2004}, DFs were used for parameterisation of the space by distances from the material features either exactly or approximately, taking into account that the defining attribute function should be at least $C^1$ continuous to avoid creases and stresses in it. In \cite{Fryazinov:2015}, an IDF based method for defining gradient materials was introduced. IDFs are represented as an approximate Euclidean shortest path and are used for interpolation between sources. In \cite{Sharma:2017}, the authors considered the decomposition of the geometry using the existing class of material distance-based functions that set up a material variation in heterogeneous objects using the medial axes transform.

\section{Distance function-based representations}
\label{sec:background}

In this section we provide some mathematical background and outline in a formalised manner four functionally-based representations that will be used to devise the hybrid function representation to be introduced in Section 4. We describe in necessary detail the mathematical basics of those representations and propose the formal definitions for two of them, namely ADF and IDF. The advantages and drawbacks of the representations are also systematically outlined.

\subsection{Mathematical background and notations}
\label{ssec:math_back_notations}

Let us introduce the mathematical definitions which will be used hereinafter. First we introduce the definition of a metric space and a distance function that follows \cite{Dyer:2014}:

\begin{definition}
    \label{def:metric_space}
    Let $X$ be a non-empty point set in a Euclidean vector space $\mathbb{R}^n$ and let function $d: X \times X \mapsto \mathbb{R}$ be such that for points $\forall \bm{p}_i \in X \subset \mathbb{R}^n$ the following conditions are satisfied:
    
    $d(\bm{p}_1,\bm{p}_2) \geq 0$; $d(\bm{p}_1,\bm{p}_2)=0 \Leftrightarrow \bm{p}_1=\bm{p}_2$; $d(\bm{p}_1, \bm{p}_2) = d(\bm{p}_2, \bm{p}_1)$; $d(\bm{p}_1,\bm{p}_2)\leq d(\bm{p}_1,\bm{p}_3)+d(\bm{p}_2,\bm{p}_3)$. Then the function $d(\bm{\cdot}, \bm{\cdot})$ is called a \textit{metric} or a \textit{distance function} on set $X$ and the pair $(X,d)$ is called a \textit{metric space}.
\end{definition}

\begin{figure}
    \centering
    \includegraphics[width=0.95\linewidth]{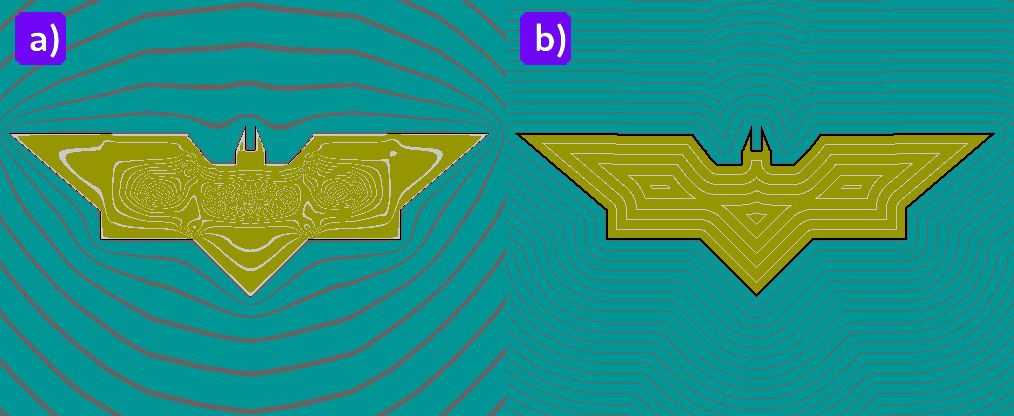}
    \caption{a) The FRep field of the functionally defined 'bat' object and b) The SDF field computed for the functionally defined 'bat' object. The colours in the pictures correspond to the point membership rule: blue colour corresponds to the negative values of the defining function, black colour corresponds to the boundary of the object and yellow colour corresponds to the positive values of the defining function.}
    \label{fig:frep_sdf}
\end{figure}

In this work we are focusing on distance-based representations for defining volumetric objects. Let us introduce a more instrumental notion for the distance function that satisfies \textbf{definition \ref{def:metric_space}} and that we will use subsequently in the next sections, as follows \cite{SCHECHTER:1997}:
\begin{definition}
    \label{def:distance_func}
    Let $X$ be a point set in a Euclidean vector space $\mathbb{R}^n$ and let $\langle \bm{\cdot}, \bm{\cdot} \rangle$ be an inner product defined in $\mathbb{R}^n$. Then the Euclidean norm of the point $\bm{p} \in X$ is defined as $||\bm{p}||=\sqrt{\langle \bm{p},\bm{p}\rangle}$. If $\bm{q}\in X$ is another point, the distance between these two points is defined as a function:
    \begin{align}
    \label{eq:dist_func}
        F_{DF}(\bm{p},\bm{q})=||\bm{p}-\bm{q}||=\sqrt{\langle \bm{p}, \bm{q} \rangle}
    \end{align}
\end{definition}

In this work we deal with functionally defined objects that are specified as closed point subsets $G \subseteq X$. As we are dealing with the objects defined by the functions, a point membership classification is used to distinguish between exterior, boundary and interior of the object. Therefore, let us introduce a formal definition of the boundary $\partial G$ of the subset $G$ as follows:
\begin{definition}
    \label{def:boundary}
    Let $G$ be a subset of the defined metric space $(X,d)$. The boundary $\partial G$ of this subset $G$ is defined as $\overline{G}\backslash G_{in}$, where $\overline{G}=\bigcap\{G_C: G_C \supseteq G \}$ is a closure of a metric space $(X,d)$, $G_C$ is a closed set in $X$, and interior of $G$ is $G_{in}=\bigcup\{G_U:G_U \subseteq G \}$, where $G_U$ is an open set in $G$.
\end{definition}

There are two important properties of the functions that we rely on in the next sections: continuity and smoothness. The continuity of the function is defined as follows \cite{Delfour:2010}:
\begin{definition}
    \label{def:continuity}
    Let $X$ be an open subset of $\mathbb{R}^n$. Let $C(X)$ be the space of continuous functions $X\mapsto \mathbb{R}^n$. Let $\mathbb{N}^n$ be the set of all tuples $\alpha=(\alpha_1, ..., \alpha_n) \in \mathbb{N}^n$. Then $|\alpha|$ is the order of $\alpha$ and $\partial^\alpha$ is the partial derivative. For an integer $k\geq 1$
    \begin{align}
        C^k(X) := \{f\in C^{k-1}(X): \partial^\alpha f \in C(X), \forall \alpha, |\alpha|=k \}
    \end{align}
    where
    \begin{align}
        |\alpha|=\sum_{i=1}^{N}\alpha_i, \quad \partial^\alpha=\frac{\partial^{|\alpha|}}{\partial x_1^{\alpha_1}...\partial x_N^{\alpha_N}}
    \end{align}
\end{definition}
In this work we discuss functions that are either at least $C^0$ or $C^1$ continuous. A function $f$ is said to be of class $C^0$ if it is continuous on $X\subset \mathbb{R}^n$. A function $f$ is said to be of class $C^1$ if it is differentiable and continuous on $X\subset \mathbb{R}^n$. 

Formally, smoothness of the function follows from the previous definition and can be defined as in \cite{SCHECHTER:1997}:
\begin{definition}
    A function $f:X \mapsto \mathbb{R}^n$ is  called smooth if it is n-times differentiable, i.e. if it belongs to a specific class of functions that can be defined as $C^n(X)$ for which $f^{(n)}$ exists and it is continuous, particularly if it satisfies $C^\infty(X)=\bigcap_{n=1}^\infty C^n(X)$.
\end{definition}

\begin{figure}
    \centering
    \includegraphics[width=0.6\linewidth]{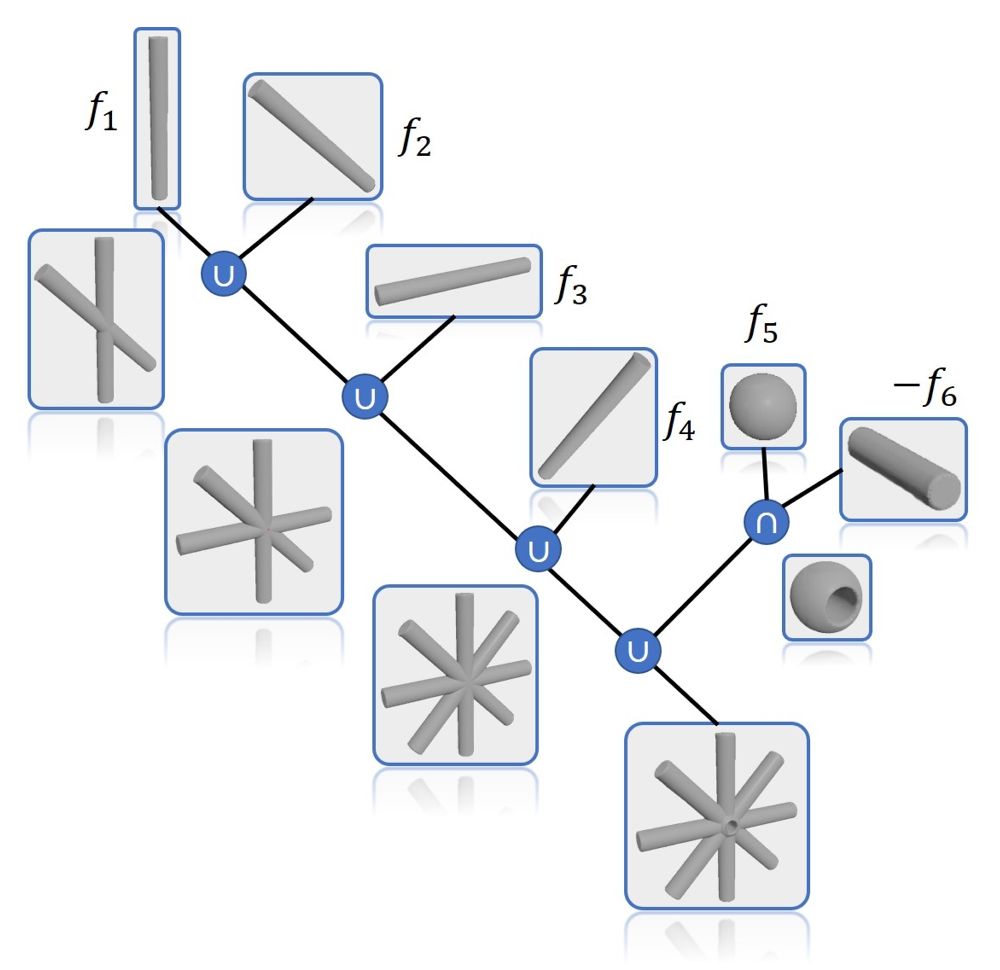}
    \caption{A constructive tree for the FRep object in the form of a 'snow flake' that was converted to SDF. This tree consists of objects defined by SDF functions $f_i$ stored in the tree leafs and operations applied to them stored in the tree nodes.}
    \label{fig:frep_tree}
\end{figure}

\subsection{Function representation}
\label{ssec:back_frep}

Let us introduce the definition of FRep \cite{Pasko:1995}:
\begin{definition}
    \label{def:frep}
    Let the geometric shape of the object $O_{FRep}$ be defined as a closed point subset $G$ of n-dimensional point set $X$ in Euclidean space $\mathbb{R}^n$ with $\bm{p}=(x_1,...,x_n) \in \mathbb{R}^n$ using a real-valued defining function $F_{FRep}(\bm{p})$. Then function representation is defined as
    \begin{align}
        \label{eq:frep_func}
        O_{FRep} := F_{FRep}(\bm{p})\geq 0
    \end{align}
\end{definition}

The FRep function (see Fig. \ref{fig:frep_sdf}, (a)) provides the information about point membership: 
\begin{align}
    \begin{cases}
    \label{eq:frep_PMembership}
        F_{FRep}(\bm{p})<0 \quad &\bm{p} \in X \backslash G 
        \\
        F_{FRep}(\bm{p})=0 \quad &\bm{p} \in \partial G
        \\
        F_{FRep}(\bm{p})>0 \quad &\bm{p}\in G_{in}
    \end{cases}
\end{align}
The major requirement for $F_{FRep}(\bm{p})$ is to be at least $C^0$ continuous. 

FRep is a high-level and uniform representation of multidimensional geometric objects. The subject of particular interest is 4D objects with fourth coordinate specified as time. FRep generalises implicit surface modelling and extends a CSG approach. FRep has a closure property as operations applied to the FRep defining functions produce continuous resulting FRep functions. The FRep object can be defined as a primitive (e.g. sphere, octahedron, cylinder, etc.) or as a complex object that is defined in the form of a constructive tree. In this case, primitives are stored in the leaves of the tree and operations are stored in its nodes. 

There exist many well-developed operations, e.g. set-theoretic operations, metamorphosis, blending and bounded blending, offsetting, bijective mapping and others \cite{Pasko:1995}.
FRep covers traditional solids \cite{Karczmarczuk:1999}, scalar fields, heterogeneous objects including both static and time dependent volumes \cite{Durikovic:2001}. 
Fig. \ref{fig:frep_sdf} (a) shows the FRep field obtained using 14 set-theoretic operations applied to triangles and rectangles to construct the 'bat'. In Fig. \ref{fig:frep_tree} we present a constructive tree that describes how a FRep object in the form of a 'snow flake', that was converted to SDF, was created using union $\cup$ and intersection $\cap$ set-theoretic operations. In general case, the FRep field is not distance-based as field isolines do not precisely follow the object shape. The advantages and drawbacks of the representation can be found in table \ref{tab:fields_comp}, in the first column.

\subsection{Signed distance function}
\label{ssec:back_sdfs}

Let us introduce the definition of SDF that relies on \textbf{definitions \ref{def:metric_space}, \ref{def:distance_func} and \ref{def:boundary}}:
\begin{definition}
    \label{def:sdf}
    Let $(X,d)$ be a metric space. Let the geometric shape G of the object $O_{SDF}$ be specified in $(X,d)$ as a point subset $G \subseteq X$. Then a signed distance function $F_{SDF}(\bm{p})$ is defined as:
    \begin{align}
    \label{eq:sdf_def}
        F_{SDF}(\bm{p}) = \begin{cases} 
                                d(\bm{p},\partial G) \quad &\text{if} \quad \bm{p} \in G \\
                                -d(\bm{p}, \partial G) \quad &\text{otherwise}
                          \end{cases}
    \end{align}
    where $d(\bm{p}, \partial G)\equiv F_{DF}(\bm{p}, \partial G)$. Then the SDF representation is defined as follows:
    \begin{align}
        \label{eq:sdf_rep}
        O_{SDF} := F_{SDF}(\bm{p}) \geq 0
    \end{align} 
\end{definition}

The SDF function is at least $C^0$ continuous as it can be not differentiable at some points of Euclidean space $\mathbb{R}^n$ and it has gradient discontinuities on the object's medial axes. The SDF representation provides the information about point membership in the same manner as FRep.

The most common operations that are defined for SDF are: offsetting \cite{Balint:2019}, surface interpolation, multiple-object averaging, spatially-weighted interpolation, texturing, blending, set-theoretic operations, metamorphosis \cite{Payne:1992} and others. SDF can be used for a material definition in heterogeneous objects \cite{BISWAS:2004}, additive manufacturing \cite{Barclay:2016}, collision detection problems, particle simulations \cite{Kim:2015} and others. 

Fig. \ref{fig:frep_sdf}, (b) shows the SDF field generated for the 'bat' object. As it can be seen, the isolines are spaced equidistantly and follow the shape of the object. The advantages and drawbacks of SDF can be seen in table \ref{tab:fields_comp}, second column.

\subsection{Adaptively sampled distance function}
\label{ssec:back_adfs}

Adaptively sampled distance function (ADF) \cite{Frisken:2000} is a distance function that is computed on hierarchical grids, e.g. tree-like data structures. ADF satisfies all the requirements of \textbf{definitions \ref{def:metric_space}, \ref{def:distance_func}} and \textbf{\ref{def:sdf}}. To our knowledge, there is no well-established formal definition of ADF in the literature. There are several works where ADF is interpreted in a different way compared to \cite{Frisken:2000}. For example, in \cite{Wang:2011} ADF was defined using T-meshes with different interpolation operation for restoring the field, in \cite{Koschier:2016} it was suggested to use a hierarchical hp-adaptation for constructing ADF, in \cite{TANG:2018} it was suggested to construct ADF using estimation of the principal curvatures of the input surface. In this work we introduce a formal definition of ADF. Let us first give the definition of the hierarchical tree structure:
\begin{definition}
    \label{def:tree}
    Let a set of nodes and edges $(Q,E)$ be an undirected connected graph $T$ that contains no loops and starts at some particular node of $T$. Then such a graph $T$ is defined as a tree.
\end{definition}

Let space $\mathbb{R}^n$ be subdivided according to the local details using some k-ary tree $T:=(Q,E)$ with nodes $q \in Q$. Each node $q$ is defined as an n-dimensional cell. According to the SDF \textbf{definition \ref{def:sdf}} we need to compute the distance to the boundary $\partial G$ of the geometric subset $G$. Taking these preliminaries into account, let us formulate the ADF definition in the constructive manner:
\begin{definition}
    \label{def:adf}
    Let the geometric shape $G \subseteq X$ of the object $O_{ADF}$ be defined in a metric space $(X,d)$. Let $(X,d)$ be subdivided into nodes $q\in Q$ with corner vertices $\bm{p}_i$ according to the level of detail using k-ary tree $T:=(Q,E)$. Let the boundary $\partial G$ be subdivided with the maximum tree depth, while $X\backslash G$ and $G_{in}$ be subdivided with some minimum tree depth. Let the corner vertices of the boundary nodes $q$ be defined as $\bm{p}_{b_i}$. Then the distance function between these points is $d(\bm{p}_i, \bm{p}_{b_i}) \equiv F_{DF}(\bm{p}_i, \bm{p}_{b_i})$. Thereafter, the ADF distance function $F_{ADF}(\bm{p})$ on the tree $T$ is restored at each node $q$ using some interpolation function $F_{I}(\bm{p})$ and is defined as follows:
    \begin{align}
    \label{eq:adf_def}
        F_{ADF}(\bm{p}) = \begin{cases} 
                                (F_I\circ F_{DF})(\bm{p}) \quad &\text{if} \quad \bm{p} \in G \\
                                -(F_I\circ F_{DF})(\bm{p}) \quad &\text{otherwise}
                          \end{cases}
    \end{align}
    The ADF representation is defined in the form of an inequation:
    \begin{align}
        \label{eq:adf_rep}
        O_{ADF}:=F_{ADF}(\bm{p})\geq 0
    \end{align}
\end{definition}

The ADF field generated as it was described in \cite{Frisken:2000} has $C^0$ discontinuities where the cells of different size appear and it has $C^1$ discontinuities caused by the bilinear/trilinear interpolation that was used for restoring a DF at each cell. The ADF representation provides the information about point membership in the same manner as FRep. The subset $X$ can be  subdivided using one of the types of k-ary trees: quadtrees or octrees. 
ADF can be used for an efficient interactive real-time modelling, e.g. sculpting, of the heterogeneous objects as the tree data structure provides fast access to object's geometry and its specified attributes. ADF is also suitable for solving surface restoration problems \cite{Deng:2008, TANG:2018}. It supports the same operations as SDF. ADF are especially suitable for dynamic simulations \cite{Koschier:2016}, for example, morphing between shapes, as a hierarchical data structure can efficiently be rebuilt at each animation frame \cite{Frisken:2000}. The advantages and drawbacks of ADF can be seen in table \ref{tab:fields_comp}, third column.

\begin{figure*}[!t]
    \centering
    \includegraphics[width=0.95\linewidth]{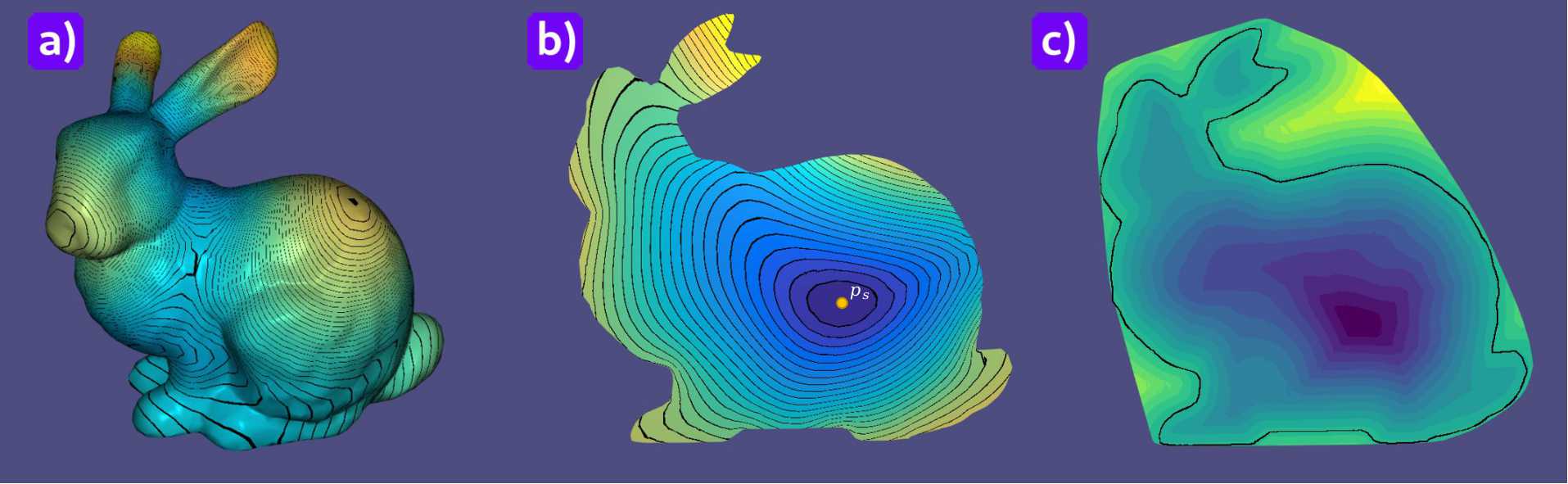}
    \caption{The IDF field computed for the 'Stanford Bunny' 3D mesh using the method described in \cite{Rustamov:2009} and an SDF slice to show the difference in nature of these fields. a) the IDF field computed on the boundary of the mesh. Black isolines show how the field is changing according to the shape of the object; b) the interior slice of the mesh with computed IDFs. The yellow point $\bm{p}_s$ in the slice corresponds to the 'source' point.
    c) the SDF slice of the same model with computed interior and exterior distances. Colour changing reflects how the distances are changing from interior to exterior of the object.}
    \label{fig:idf}
\end{figure*}

\subsection{Interior distance function}
\label{ssec:back_idfs}

Interior distance function (IDF) is not a well-established notion yet as in literature there is neither a general approach for generating DFs of this rather broad nature nor one unique name for them. In this work we suggest to use this notion for a representation with a defining function obtained as follows: the distance function is computed on the boundary of the object and then the generated distances are smoothly interpolated in its interior. Let us introduce the definition of IDF that relies on the definitions specified in subsection \ref{ssec:math_back_notations}: 
\begin{definition}
    \label{def:idf}
    Let the geometric shape $G \subseteq X$ of the object $O_{IDF}$ be defined in a metric space $(X,d)$. Let points $\bm{p}_{b_i}$ belong to $\partial G$, and let points $\bm{p}_{in_k}$ belong to $G_{in}$. Let a distance function $d(\bm{p}_{b_i},\bm{p}_{b_j})\equiv F_{DF}(\bm{p}_{b_i},\bm{p}_{b_j})=||\bm{p}_{b_i}-\bm{p}_{b_j}||_{\mathbb{R}^n}$ between any boundary points $\bm{p}_{b_i}$ and $\bm{p}_{b_j}$ on a curved domain $\partial G$ be recovered. Thereafter, by constructing an interpolation function $F_I(F_{DF}(\bm{p}_{b_i}, \bm{p}_{b_j}), \bm{p}_{in_k})$ that is at least $C^1$ continuous, boundary distances are extended to interior of the object $O_{IDF}$. Therefore, the IDF function can be defined as:
    \begin{align}
        \label{eq:idf}
        F_{IDF}(\bm{p}_{in_k}) = F_I(F_{DF}(\bm{p_{b_i}}, \bm{p_{b_j}}), \bm{p}_{in_k})
    \end{align}
    where $ 0 \leq i,j, < N$, $N$ is the number of boundary points, $0\leq k < M$, $M$ is the number of interior points. The IDF representation is defined in the form of an inequation:
    \begin{align}
        \label{eq:idf_rep}
        O_{IDF}:= F_{IDF}(\bm{p}) \geq 0
    \end{align}
\end{definition}

IDF is usually obtained by solving a partial differential equation (PDE) or applying some numerical method, e.g., graph approaches \cite{Liu:2011} or Markov chains \cite{COIFMAN:2006}. 
Among PDE-based methods the following methods can be considered as representative: geodesic distances obtained as the solution of heat equation \cite{Crane:2013}, diffusion maps combined with smooth barycentric interpolation of the distances in interior of the object \cite{Rustamov:2009}, the optimal mass transport \cite{Solomon:2014} and some others. IDF is usually used in the tasks related to shape analysis \cite{Rustamov:2009}, geometry restoration \cite{Patane:2012}, morphing and less commonly for an attribute definition in interior of the object \cite{Fryazinov:2015}. The advantages and drawbacks of IDF can be found in table \ref{tab:fields_comp}, last column.

In Fig. \ref{fig:idf} we show how the approach described in \cite{Rustamov:2009} can be applied to the polygonal mesh of the 'Stanford Bunny'. The distances are computed on the boundary, as it can be seen in Fig. \ref{fig:idf} (a), using the diffusion maps, and then propagated in interior of the object, as it can be seen in Fig. \ref{fig:idf} (b), using the barycentric interpolation. For convenience of data visualisation we compute distances from the fixed 'source' point $\bm{p}_s$ to other points of the mesh. If we compare two pictures shown in Fig. \ref{fig:idf} (b) and (c), we can see that the distance fields obtained in interior of the bunny are completely different. The IDF field (b) is smooth and continuous while the SDF field (c) is not smooth and has some sharp features in interior of the object. 

\begin{table}[!t]
    \scriptsize
    \centering \resizebox{\textwidth}{!}{
    \begin{tabular}{|c|c|c|c|c|}
    \hline
    & FRep & SDF & ADF & IDF \\
    \hline
    \rotatebox[origin=c]{90}{advantages} &
    \multicolumn{1}{m{3.5cm}|}{
    \begin{itemize}[leftmargin=*]
        \item FRep generalises implicit surface modelling and extends a constructing modelling approach;
        \item FRep supports point membership;
        \item FRep is closed guaranteeing to get an at least $C^0$ continuous resulting function;
        \item FRep covers solids, scalar fields, volumes, time-dependent volumes and hypervolumes for heterogeneous object modelling.
        \item FRep has many well-developed operations that support multidimensional transformations in $\mathbb{R}^n$; 
    \end{itemize}
    } & 
    
    \multicolumn{1}{m{3.5cm}|}{ 
    \begin{itemize}[leftmargin=*]
        \item SDF provides distances to the object surface both inside and outside it; 
        \item SDF defines a watertight object; 
        \item SDF is a Lipschitz continuous function; 
        \item SDF is Fre\'chet differentiable almost everywhere;
        \item SDF satisfies the solution of the eikonal equation; 
        \item SDF supports point membership; 
        \item SDF is effectively discretised, has a predictable field behaviour and is efficiently rendered.
    \end{itemize}
    } &  
    
    \multicolumn{1}{m{3.5cm}|}{
    \begin{itemize}[leftmargin=*]
        \item ADF data structure efficiently subdivides the Euclidean space $\mathbb{R}^n$ according to the level of detail; 
        \item ADF distances are adaptively sampled; 
        \item ADF supports point membership;  
        \item ADF possesses an efficient memory management: in a small amount of memory a significant amount of information about the object can be stored; 
        \item ADF hierarchical tree data structure is fast to rebuild that makes it possible to handle time-variant objects; 
        \item ADF can be efficiently rendered in real time.
    \end{itemize}
    } &
    
    \multicolumn{1}{m{3.5cm}|}{
    \begin{itemize}[leftmargin=*]
       \item IDF is %at least $C^1$ continuous and
       shape-aware; 
       \item IDF is deformed with the boundary; 
       \item IDF is smooth; 
       %\item IDF is not affected by the boundary noise;
       \item IDF is suitable for the distance-based attribute definition in interior of the object.
    \end{itemize}
    }
    \\
    \hline
    \rotatebox[origin=c]{90}{drawbacks} & 
    \multicolumn{1}{m{3.5cm}|}{ 
    \begin{itemize}[leftmargin=*]
        \item Distances can be obtained for a limited number of FRep objects;
        \item FRep object can have a boundary with dangling portions that are not adjacent to the interior of the object;
        \item FRep has an unpredictable non-distance based behaviour of the resulting field and, as a consequence, it is sometimes problematic to render in 3D.
    \end{itemize}
    } & 
 
    \multicolumn{1}{m{3.5cm}|}{
    \begin{itemize}[leftmargin=*]
        \item SDF is not differentiable at some points of Euclidean $\mathbb{R}^n$ space. Loss of SDF differentiability happens when the current point is sufficiently close to a concave singularity (a concave corner/edge); 
        \item SDF has discontinuous gradients on the object's medial axes; 
        \item SDF is not smooth;
        \item SDF is not suitable for attribute modelling due to $C^1$ discontinuity.
    \end{itemize}
    } &  
 
    \multicolumn{1}{m{3.5cm}|}{
    \begin{itemize}[leftmargin=*]
        \item ADF field has $C^0$ discontinuities where cells of the different size appear as the result of the hierarchical subdivision; 
        \item ADF field has $C^1$ discontinuities that are introduced by the bilinear/trilinear interpolation \cite{Frisken:2000} during reconstruction of the field at each cell; 
        \item ADF is not suitable for attribute modelling due to $C^0$ and $C^1$ discontinuities.
    \end{itemize}
    } &
 
    \multicolumn{1}{m{3.5cm}|}{
    \begin{itemize}[leftmargin=*]
       \item IDF can be computationally expensive; 
       \item IDF field accuracy for some methods is highly dependent on a time step and type of the used discretisation; 
       \item IDF is defined only in interior of the object.
    \end{itemize}
    }
    \\ 
    \hline
    \end{tabular}
    }
    \caption{Comparison table of the advantages and drawbacks of FRep, SDFs, ADFs and IDFs.}
    \label{tab:fields_comp}
\end{table}

\subsection{Heterogeneous objects}
\label{ssec:heter_objs}

In the previous subsections we have discussed how geometric shape of objects can be defined using distance-based methods. In this subsection we discuss how attributes can be considered in concert with the geometric shape of the object to represent the heterogeneous object. Let us first introduce a general definition of the heterogeneous object.
\begin{definition}
    \label{def:heter_obj}
    Let the object $O_H$ be defined as a two component tuple: geometric shape $G \subseteq X$ in the form of a multidimensional point-set geometry and attributes $A_i$ corresponding to the physical properties of the object $O_H$.
    Then such object $O_H$ is a heterogeneous object defined as:
    \begin{align}
    \label{eq:heter_obj}
        O_H:=(G, A_1, ..., A_n),
    \end{align}
    where $n \in \mathbb{N}$ is the number of attributes.
\end{definition}

Attribute distributions specified in heterogeneous objects $O_H$ can be uniform or non-uniform. For instance, the simple example of the uniform distribution can be a homogeneously coloured object. As to non-uniformity, it can be presented as porous structures or microstructures with non-linear varying  density.

In this work we will apply the hypervolume model \cite{Pasko:2001} to define heterogeneous objects $O_{H_V}$ using FRep or any other distance function-based representation. A hypervolume object is defined as follows:

\begin{definition}
    \label{def:hypervol}
    Let the geometric shape G of $O_{H_V}$ be defined by a real-valued function $F_G(\bm{p}), \bm{p} \in \mathbb{R}^n$ that is at least $C^0$ continuous and let attributes be defined by any $F_{A_i}(\bm{p})$. Then heterogeneous object $O_{H_V}$ is defined as:
    \begin{align}
        \label{eq:hypervol_object_def}
        O_{H_V} := (G,A_1,...,A_n):(F_G(\bm{p}), F_{A_1}(\bm{p}),...,F_{A_n}(\bm{p})),
    \end{align}
    where $n \in \mathbb{N}$ is the number of attributes.
\end{definition}

In general case, attribute functions $F_{A_i}(\bm{p})$ are not necessarily continuous. However, as it was shown in \cite{BISWAS:2004}, better control of the attributes on the surface and in the interior of the distance-based objects can be achieved when the attribute defining functions are parameterised by the distances.
The main requirement for the distance function is to be at least $C^1$ continuous. This requirement prevents the appearing of stress concentrations, creases and other singularities in modelled attribute distributions.

There are several interesting examples discussed in \cite{BISWAS:2004}. In particular, the distance-based smooth and differentiable attribute functions were applied to represent a parabolic distribution of the graded refractive index in Y-shaped solid of the waveguide. In this case, it is important that the distribution of the index of refraction is uniform and smooth. Another example is to use such distance-based attribute functions for modelling different types of materials, e.g. silicon carbide (SiC). It is important to note that the approach introduced in \cite{BISWAS:2004} was not applied to such attributes as textures, colours and similar attributes.

\begin{figure*}[!t]
    \centering 
    \includegraphics[width=1.0\linewidth]{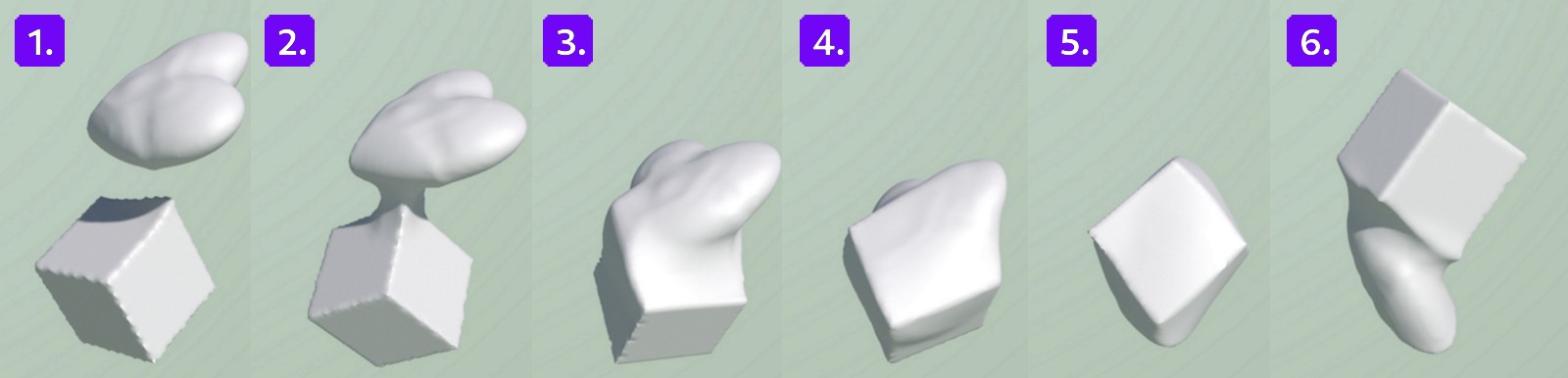}
    \caption{The STB-based metamorphosis operation over the initially FRep 'heart' converted to HFRep and initially BRep 'cube' converted to SDF 'cube'.}
    \label{fig:stb_example}
\end{figure*}

\section{Hybrid function representation}
\label{sec:hfrep_theory}

In this section we introduce and systematically describe a general approach for defining heterogeneous volumetric objects using a hybrid function representation (HFRep). First, we list the requirement to HFRep, then outline its mathematical basics, and finally describe its properties with respect to four basic DF-based representations. 

\subsection{Problem statement}
\label{ssec:hfrep_problem_st}

Let us give the exact problem statement. Our goal is to propose a hybrid function representation (HFRep) that is suitable for defining volumetric heterogeneous objects. We assume that the geometric shape $G$ of the given object is defined by FRep, and its defining function is known. To devise the HFRep embracing advantages and circumventing disadvantages of FRep, SDF, ADF, IDF, it is essential to  obtain a real-valued defining function in an n-dimensional Euclidean space with the following properties:

\begin{enumerate}
    \item the HFRep function should provide sufficiently accurate distance approximation in Euclidean space $\mathbb{R}^n$ without $C^0$ and $C^1$ discontinuities. 
    \item the HFRep function should be at least $C^0$ continuous with possibility to enforce it to be at least $C^1$ continuous.
    \item the HFRep function should satisfy the point membership test: it should be positive in interior of the geometric shape $G$, take exact zero values only at the object boundary $\partial G$ and it should be negative in exterior of the geometric shape $X \backslash G$;
    \item the HFRep should be a multidimensional object representation; in particular, dealing with 4D objects is of paramount importance to cover time-variant models with the fourth 'time' coordinate;
    \item the HFRep representation should be suitable for the heterogeneous object modelling allowing for defining attribute functions related to the geometry;
    \item the HFRep attribute functions should depend on evaluation point $\bm{p} \in G$ and be parameterised by distance values of the obtained HFRep geometry function.
\end{enumerate}

The fulfilment of these conditions guarantees that the generated object will be watertight and such operations as blending and metamorphosis will not suffer from creases. Overall, the defining HFRep function that is considered in concert with attribute functions parameterised by distances will be suitable for dealing with multi-material aspects of heterogeneous objects including time-variant ones.

\begin{figure} 
    \centering 
    \includegraphics[width=0.5\linewidth]{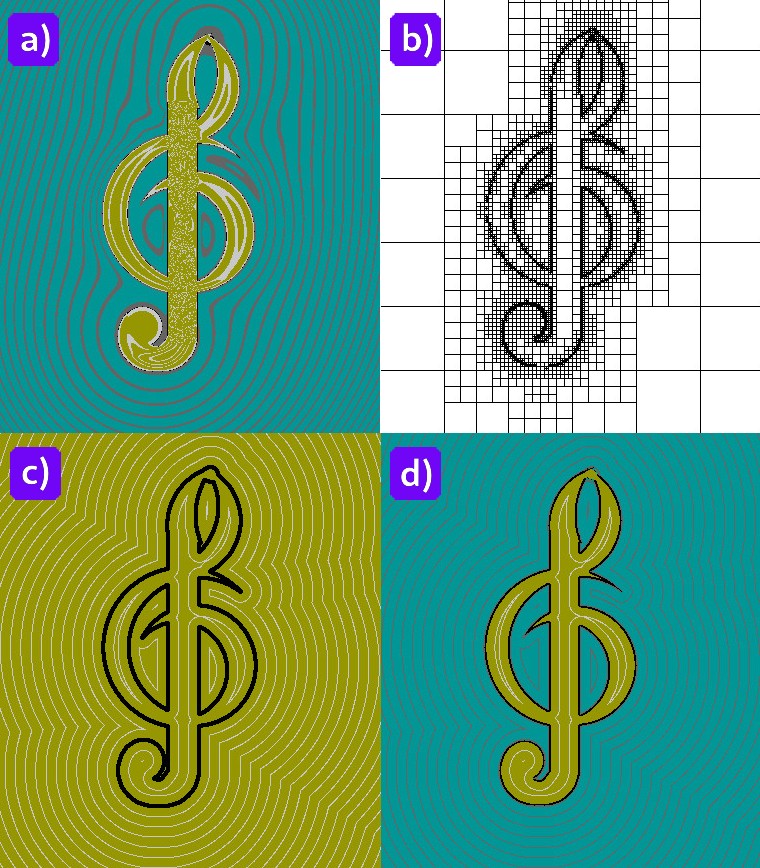}
    \caption{The illustration of HFRep based on FRep and ADF with applied PHT-spline (a polynomial spline over hierarchical T-mesh) interpolation to restore the distance field at each cell. ADFs are generated using a numerical solution of the eikonal equation on the quadtree. a) the FRep field; b) a hierarchical quadtree subdivision; c) UDF computed on the quadtree with the applied PHT-spline interpolation for restoring distances at each quadtree cell; d) the HFRep field that was obtained using the generated ADF.}
    \label{fig:hfrep_adf_example}
\end{figure}

\subsection{Definition of the hybrid function representation}
\label{ssec:hfrep_definition}

First, we provide a mathematical definition for the geometric aspects of HFRep. Then we add the part related to attributes. The geometric shape $G$ of an HFRep object $O_{HFRep}$ is defined as follows:
\begin{definition}
    \label{def:hfrep_geom}
    Let the geometric shape $G \subseteq X$ of the object $O_{HFRep}$ be defined in a metric space $(X,d)$. Given at least $C^0$ or $C^1$ continuous FRep function $F_{FRep}(\bm{p})$, the distance to the object boundary $\partial G$ is defined as $(F_I\circ F_{DF})(\bm{p}, \partial G) \equiv (F_I\circ F_{DF})(\bm{p})$, where $F_I(\bm{\cdot})$ is at least $C^1$ continuous interpolation function and $d(\bm{\cdot}, \bm{\cdot})\equiv F_{DF}(\bm{\cdot}, \bm{\cdot})$ is a distance-based function, in particular SDF, ADF or IDF. Then the HFRep function is defined as follows:
    \begin{align}
        \label{eq:hfrep_func}
        F_{HFRep}(\bm{p}) = (F_{sign}\circ F_{FRep})(\bm{p})\cdot (F_I \circ F_{DF})(\bm{p})
    \end{align}
    where $F_{sign}(\bm{\cdot})$ is an at least $C^1$ continuous function that provides a sign for the computed function $(F_I \circ d)(\bm{p})$ and satisfies the FRep point membership test, equation (\ref{eq:frep_PMembership}). Finally, the HFRep representation is defined as:
    \begin{align}
    \label{eq:hfrep_rep}
        O_{HFRep} := F_{HFRep}(\bm{p})\geq 0
    \end{align}
\end{definition}

The continuity of the HFRep function $F_{HFRep}(\bm{p})$ depends on the continuity of the FRep function $F_{FRep}(\bm{p})$. In the case when we are dealing only with geometric shapes, it is sufficient to have $C^0$ continuity for the HFRep function. Otherwise, in case of heterogeneous object modelling, the HFRep function should belong to the class of functions that are at least $C^1$ continuous. We will give details on how to control the continuity of the HFRep function later in this section.

\begin{table*}[!t]
    \scriptsize
    \centering \resizebox{\textwidth}{!}{
    \begin{tabular}{|c|c|c|c|}
    \hline
    Inherited from FRep &Inherited from SDF &Inherited from ADF &Inherited from IDF \\
    \hline
    \multicolumn{1}{|m{4cm}|}{
    \begin{itemize}[leftmargin=*]
        \item The continuity of the HFRep function depends on the continuity of the FRep function. 
        \item The HFRep object is watertight. 
        \item HFRep represents  multidimensional objects, in particular 4D objects with the fourth coordinate specified as time.
    \end{itemize}
    }
    &
    \multicolumn{1}{m{4cm}|}{
    \begin{itemize}[leftmargin=*]
        \item HFRep provides at least $C^0$ continuous distance function.
        \item the HFRep object is watertight. 
        \item the HFRep function is Lipshitz continuous and Fre\'chet differentiable everywhere; 
        \item the HFRep function satisfies the solution of the eikonal equation; 
        \item the HFRep object can be efficiently discretised and rendered.
    \end{itemize}
    }
    &
    \multicolumn{1}{m{4cm}|}{
    \begin{itemize}[leftmargin=*]
        \item HFRep provides at least $C^0$ continuous distance function for any FRep object that was spatially subdivided according to the local details using a hierarchical data structure. 
        \item Hierarchical data structure can also be  used for defining and storing object's attributes. 
    \end{itemize}
    }
    &
    \multicolumn{1}{m{4cm}|}{
    \begin{itemize}[leftmargin=*]
        \item HFRep provides at least a $C^0$ continuous unsigned distance function for any FRep object in its interior if IDF is used for obtaining distances; 
        \item Distances in the interior of the HFRep object are shape-aware, deformed with boundaries and are not affected by the boundary noise. 
        \item There is also a potential for modelling attributes in interior of the volumetric object. 
    \end{itemize}
    }
    \\
    \hline
    \end{tabular}
    }
    \caption{Properties of the hybrid function representation that depend on the combination of FRep with one of the distance fields.}
    \label{tab:hfrep_props}
\end{table*}

Now let us show that $F_{HFRep}(\bm{p})$ continuity is either $C^0$ or $C^1$. By applying a smoothing interpolation function $F_I(\bm{\cdot})$ that is at least $C^1$ continuous to the discrete unsigned distance field (UDF) obtained using $F_{DF}(\bm{p}, \partial G) \in C^0$, we enforce the property $(F_I\circ F_{DF})(\bm{p})\in C^1$. The composition of functions $(F_{sign}\circ F_{FRep})(\bm{p})$ is at least $C^0$ or $C^1$ continuous, depending on the continuity of $F_{FRep}(\bm{p})$. The theorem about the continuity of the composition of two continuous functions was proofed in \cite{Boman:1967}. Therefore the continuity of the HFRep function is defined as: $C_{HFRep}=\min(C^m_{F_{sign}\circ F_{FRep}}, C^k_{F_{I}\circ F_{DF}})$, where $m = 0$ or $m=1$, $k=1$, i.e. the minimum class of continuity between two function compositions.

Now on the basis of \textbf{definition \ref{def:hypervol}}, we can formulate the definition of the heterogeneous HFRep object $O_{H_{V, HFRep}}$ as follows: 
\begin{definition}
    \label{def:hfrep_attr}
    Let the geometric shape $G$ of $O_{H_{V, HFRep}}$ be defined by at least $C^1$ continuous $F_{G}(\bm{p}) = F_{HFRep}(\bm{p})$ distance-based function. Let the attribute $A_i$ be defined as a real-valued function $F_{A_i}(F_{HFRep}(\bm{p}),\bm{p})$. Then the HFRep heterogeneous object $O_{H_{V, HFRep}}$ is defined as: 
    \begin{align}
        \label{eq:hfrep_hypervolume}
        O_{H_{V, HFRep}} := \begin{cases}
                            F_G(\bm{p}) := F_{HFRep} \geq 0 \\
                            F_{A_i}(F_{HFRep}(\bm{p}), \bm{p}), \quad i=[0,..,n] \in \mathbb{N}
                            \end{cases}
        \end{align}
        where $n$ is a number of attributes.
\end{definition}

The properties of the introduced hybrid function representation are outlined in table \ref{tab:hfrep_props}. For a particular combination of FRep with one of the distance fields, namely SDF, ADF or IDF, only one type of properties can be inherited. We show some particular properties, mentioned in the table \ref{tab:hfrep_props}, using several examples that will be discussed further in this subsection.

Fig. \ref{fig:stb_example} shows a metamorphosis between two  oscillating 4D geometric shapes ('heart', initially the FRep object then converted to HFrep; 'cube', initially the BRep object then converted to SDF) using the space-time blending (STB) method \cite{Tereshin_EG:2020}. The result is a non-distance functionally defined watertight object that is continuous and smooth.

In Fig. \ref{fig:hfrep_adf_example} (d), we demonstrate the restored distance field computed on the hierarchical grid obtained for the initial FRep object  defined as a 'treble clef', Fig. \ref{fig:hfrep_adf_example} (a) . There is neither $C^0$ nor $C^1$ discontinuities in the field as it can be seen in Fig. \ref{fig:hfrep_adf_example}, (c) or (d). All the isolines are smooth and continuous.   

\begin{figure} 
    \centering 
    \includegraphics[width=0.95\linewidth]{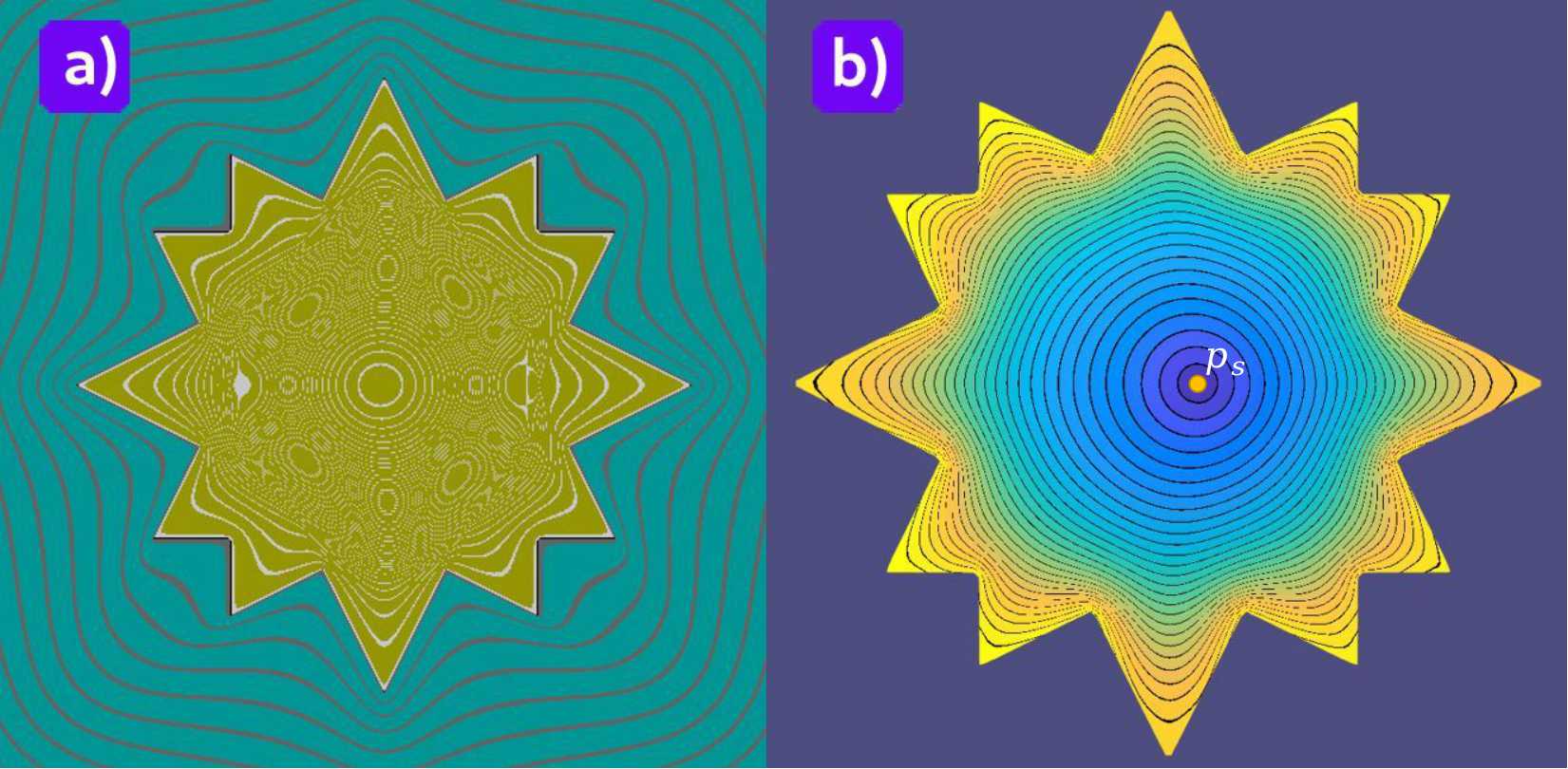}
    \caption{(a) 'Star' object and its FRep field; (b) the HFRep 'star' object generated on the basis of the FRep object. The boundary of the FRep object (a) was extracted and then used for computing boundary distances. The obtained distances were interpolated in interior of the HFRep 'star' object using barycentric interpolation and mean-value coordinates. The isolines and colour show how the field changes from the source point (white circle) towards the object boundary.}
    \label{fig:hfrep_idf_example}
\end{figure}

In Fig. \ref{fig:hfrep_idf_example}, (b) we show a simple example of interior distances computed for the FRep 'star' object, Fig. \ref{fig:hfrep_idf_example} (a), that was constructed using seven set-theoretic operations. First the boundary of the FRep object was extracted for computing boundary distances. Then the interior of the obtained convex contour was triangulated. Finally, the boundary distances were propagated in interior of the shape as it is described in \cite{Rustamov:2009}. The black isolines show that the obtained field is at least $C^1$ continuous as they are smoothly changing in the object interior.

\section{Basic algorithm for generating HFRep}
\label{sec:hfrep_basic_alg}

Let us outline in a step-by-step manner the algorithmic solution on generating the HFRep functions. The basic algorithm covers all paired combinations of FRep with DF representations, namely SDF, ADF and IDF, and allows to generate both a geometric shape and attributes. Some steps of the basic algorithm will be slightly different depending on the particular type of the DF paired with FRep. Let us start from the algorithm for generating a geometric shape of the object $O_{H_{V,HFRep}}$. Fig. \ref{fig:field_sequence} demonstrates the generated function field for each step of the basic algorithm.

\subsection{Algorithm for HFRep geometry generation}
\label{ssec:alg_hfrep_geo}
\begin{enumerate}
    \item According to the \textbf{definition \ref{def:hfrep_geom}}, we start the construction of an HFRep object $O_{HFRep}$ from defining the FRep function $F_{FRep}(\bm{p})$ for its geometric shape $G$. The FRep function $F_{FRep}(\bm{p})$ can be defined analytically, with function evaluating algorithm or using a point cloud for which it is possible to obtain a real-valued at least $C^0$ continuous $F_{FRep}(\bm{p})$. It could also be a a complex FRep object that is obtained in the form of a constructive tree.
    
    At this step we can also enforce HFRep function $F_{HFRep}(\bm{p})$ to be at least $C^1$ continuous as its continuity depends on the continuity of $F_{FRep}(\bm{p})$. We have to examine the obtained $F_{FRep}(\bm{p})$ for continuity and differentiability. The most practically used FRep set-theoretic operations in the form of the following R-function system are \cite{Pasko:1995}:
    \begin{align}
        \label{eq:R_sys_frep}
        f_{\cup}(f_1(\bm{p}), f_2(\bm{p})) &= f_1 + f_2 + \sqrt{f_1^2 + f_2^2}
        \\
        f_{\cap}(f_1(\bm{p}), f_2(\bm{p})) &= f_1 + f_2 - \sqrt{f_1^2+f_2^2} \nonumber
    \end{align}
    These functions have $C^1$ discontinuity in points where both arguments are equal to zero. Accordingly, the resulting function will only be $C^0$ continuous. If we need to obtain an at least $C^1$ continuous resulting function, we can apply another R-function system that is at least $C^{n-1}$ continuous \cite{Rvachev:1982}:
    \begingroup\makeatletter\def\f@size{7.5}\check@mathfonts
    \def\maketag@@@#1{\hbox{\m@th\large\normalfont#1}}%
    \begin{align}
        \label{eq:R0_sys}
        f_{\cup}(f_1(\bm{p}), f_2(\bm{p})) &= \begin{cases}
        f_1f_2(f_1^n+f_2^n)^{-\frac{1}{n}}, \quad &\forall f_1 > 0, f_2 > 0;  \\
        f_1, \quad &\forall f_1\leq 0, f_2 \geq 0;  \\
        f_2, \quad &\forall f_1\geq 0, f_2 \leq 0;  \\
        (-1)^{n+1}(f_1^n + f_2^n)^{\frac{1}{n}}, \quad &\forall f_1 < 0; f_2 < 0;            \end{cases}
    \end{align}
    
    \begin{align}
        f_{\cap}(f_1(\bm{p}), f_2(\bm{p})) &=\begin{cases}
        (f_1^n+f_2^n)^{\frac{1}{n}}, \; &\forall f_1 > 0, f_2 > 0; \\
        f_2, \; &\forall f_1\leq 0, f_2 \geq 0; \\
        f_1, \; &\forall f_1\geq 0, f_2 \leq 0; \\
        (-1)^{n+1}f_1f_2(f_1^n + f_2^n)^{-\frac{1}{n}}, \; &\forall f_1 < 0; f_2 < 0; 
                                          \end{cases} 
        \nonumber
    \end{align}
    \endgroup
    where $f_1(\bm{p})$ and $f_2(\bm{p})$ are FRep functions.
    
    Fig. \ref{fig:field_sequence}, (a) shows the FRep field obtained for the 'robot' object, that was generated using 39 set-theoretic operations, equation (\ref{eq:R_sys_frep}), applied to circles and rectangles.
    \begin{figure}[!t]
        \centering 
        \includegraphics[width=0.95\linewidth]{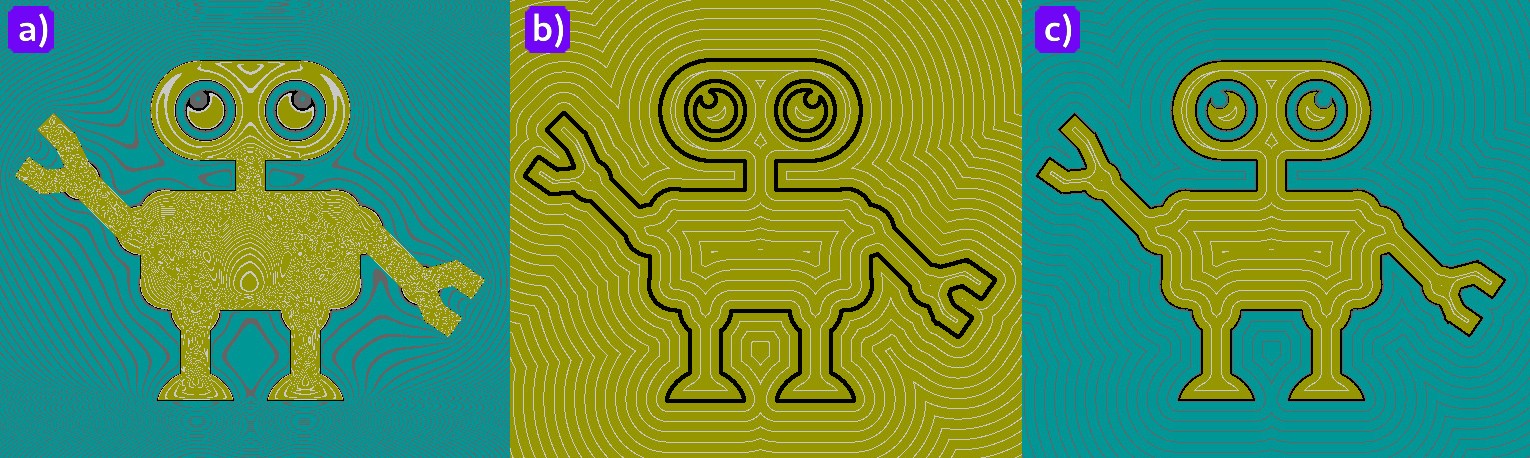}
        \caption{The illustration of the basic algorithm: a) step 1: the computed field of the 'robot' FRep object; b) steps 2 - 3: the computed unsigned distance field that can be obtained using, e.g., the distance transform or a numerical solution of the eikonal equation. The obtained field is smoothed using some spline interpolation; c) step 4: the generated HFRep field.}
        \label{fig:field_sequence}
    \end{figure}
    \newline
    \item The values of the function  $F_{FRep}(\bm{p})$ are used as an input for computing distance functions $F_{DF}(\bm{p}, \partial G)$ that should satisfy one of the \textbf{definitions} \textbf{\ref{def:sdf}, \ref{def:adf}} or  \textbf{\ref{def:idf}}. At this step we obtain an unsigned distance function that is defined as:
    \begin{align}
        \label{eq:basic_df}
        F_{DF}(\bm{p}) = d(\bm{p}, \partial G), \quad \forall \bm{p} \in X
    \end{align}
    Fig. \ref{fig:field_sequence}, (b) shows the unsigned distance field that was obtained on the basis of a typical SDF generation algorithm \cite{LEYMARI:1992}. 
    
    If the distances are computed using ADF, first, we need to subdivide the space using a hierarchical data-structure, e.g. quadtree, Fig. \ref{fig:hfrep_adf_example}, (b) and during it's construction we also need to compute basis functions, basis vertices and extraction operators for the hierarchical splines. Then we need to compute the distances at the corner vertices of each cell. Finally, we restore distances in interior of each cell using at least $C^1$ continuous spline-based interpolation to obtain a smooth and continuous distance field, e.g. shown in Fig. \ref{fig:hfrep_adf_example}, (c).
    
    Specifically for IDFs, the function $F_{DF}(\bm{p})$ is defined according to equation (\ref{eq:idf}). Distances are computed on the boundary of the object $O_{FRep}$ and then interpolated in its interior. In Fig. \ref{fig:hfrep_idf_example} (a), we can see the field of the FRep-defined 'star' object that was used for generating HFRep IDF-based field that is shown in Fig. \ref{fig:hfrep_idf_example}, (b).
    
    In Fig. \ref{fig:extrapolation_idea}, (a) we show a possible extrapolation scheme that can be used to obtain distances in exterior of the object and make an IDF-based field signed at the last step of this algorithm. To do this, we need to use the boundary distances (Fig. \ref{fig:extrapolation_idea}, (a), dark blue circles) and an appropriate at least $C^1$ continuous extrapolation operation, that will be used for obtaining distances outside the object (\ref{fig:extrapolation_idea}, Fig.7(a), red circles).
\newline
    \item The distance field obtained at the previous step is unsigned and discrete as it was computed on the finite point subset $X\subset \mathbb{R}^n$. To enforce the continuity and smoothness of the computed field, we need to apply some at least $C^1$ continuous interpolation function $(F_{I}\circ F_{DF})(\bm{p})$ to the generated unsigned field, e.g. spline-based:
    \begin{align}
        \label{eq:basic_inter}
        F_{smDF}(\bm{p}) = (F_{I}\circ F_{DF})(\bm{p})
    \end{align}
    We also need to apply a smoothing operation to an IDF field if at the previous step an extrapolation operation was applied. Otherwise, IDFs are smooth as smoothness is their inherent property. An important requirement for the interpolation function $F_I(\bm{\cdot})$ is to avoid introducing extra zeros in the distance field generated using function $F_{DF}(\bm{p}, \partial G)$.   
    \newline
    \item Finally, as the distance field obtained after previous steps is unsigned, we need to restore the field sign to distinguish between exterior $X \backslash G$, boundary $\partial G$ and interior $G_{in}$ of the object $O_{H_{V,HFRep}}$. We suggest to use some at least $C^1$ continuous step-function $F_{st}(F_{FRep}(\bm{p}))$ with the scope $[-1,1]$, that depends on the values of the defining FRep function $F_{FRep}(\bm{p})$ and approximates its well-defined behaviour ($-1$ in exterior of the object, $0$ on the boundary of the object and $+1$ in interior of the object). Therefore, the resulting HFRep function $F_{HFRep}(\bm{p})$ is defined according to \textbf{definition \ref{def:hfrep_geom}} as follows:
    \begin{align}
        \label{eq:basic_hfrep}
        F_{HFRep}(\bm{p})=(F_{st}\circ F_{FRep})(\bm{p})\cdot F_{smDF}(\bm{p})
    \end{align}
    The HFRep field generated by this function can be seen in Fig. \ref{fig:field_sequence}, (c). After a geometric shape of the HFRep object $O_{HFRep}$ was generated, we can apply different operations to it provided that they are realised by functions which are at least $C^0$ continuous. The HFRep object is also compatible with other distance-based objects. However, to preserve the distance properties for the object obtained after applying multiple operations, we might need to apply the steps of this algorithm again to this object.
\end{enumerate}

\begin{figure}
    \centering 
    \includegraphics[width=0.95\linewidth]{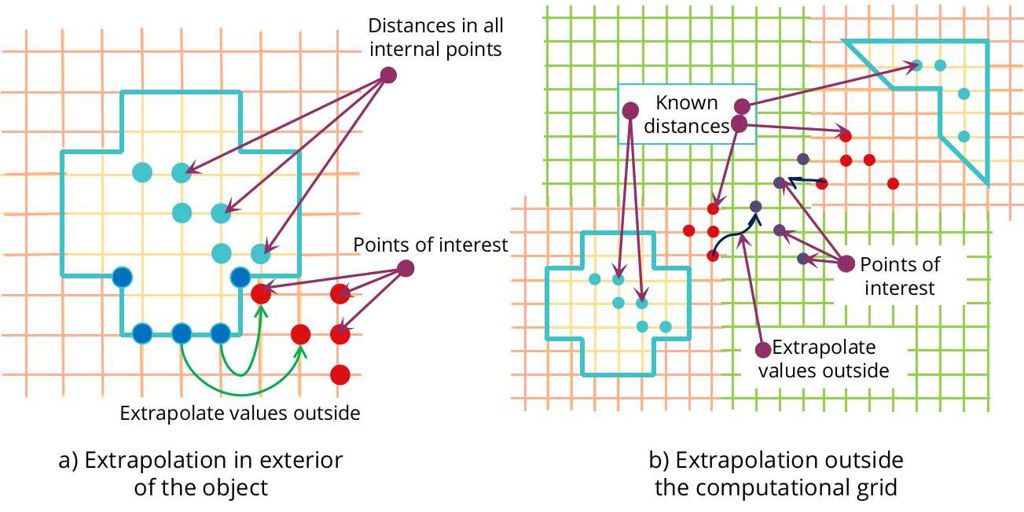}
    \caption{Two cases when extrapolation is important to enforce the continuity of the field: a) when we have DF (light pink and light orange colours) for two objects that are distantly placed in space. In this case we need to extrapolate the distance values into the points of the green grid; b) when we computed the IDF and it is essential to obtain distances in exterior of the object.}
    \label{fig:extrapolation_idea}
\end{figure}

There is a limited number of operations that preserve the distance property for the geometric shape $G$ of the object obtained after their application. These operations are rigid (Euclidean) transformations: rotations, translations, reflections or their combination. Another distance preserving operations \cite{Payne:1992} are affine translations, offsetting, linear surface interpolation, surface blurring and compression, set-theoretic operations in the form of $\min(f_1(\bm{p}), f_2(\bm{p})$ or $\max(f_1(\bm{p}), f_2(\bm{p})$ \cite{Rvachev:1982}.

In cases of other operations \cite{Reiner:2011} (e.g., scaling, blending, space-time blending, twisting, tapering and sweeping, set-theoretic operations in the form of R-functions \cite{Pasko:1995}) after their application, we have to apply the  basic algorithm to the obtained object to restore the distance property.

To make the HFRep representation continuous on the whole domain of the Euclidean space $\mathbb{R}^n$, we suggest to apply some at least $C^1$ continuous extrapolation operation to the generated field of the object. To explain this idea in more details let us consider the following example shown in Fig. \ref{fig:extrapolation_idea}, (b). In this figure we have two blue objects defined on their own pink grids and spaced from each other, so their defining grids are not overlapping. If we want to work with them, e.g. by applying some operation, we need somehow to define the distances in the points of interest of the green grid. One can extrapolate and average the distances between two pink grids and avoid full reinitialisation of the distances for both objects.

\subsection{Algorithm for HFRep attribute definition}
\label{ssec:alg_hfrep_attr}

To set up the attributes in interior of the HFRep object $O_{H_{V,HFRep}}$, we assume that we have obtained a $C^1$ continuous distance function for a geometric shape. Now we can deal with the attributes that are parameterised by the distances as it was required by  \textbf{definition \ref{def:hfrep_attr}}. Object attributes could be of different nature and there is no single algorithm to define all of them. In this work we consider such attributes as colours, microstructures, and simple 2D and volumetric textures based on noise functions parameterised by distances. 

Let us formulate the basic algorithm for specifying an attribute component $A_i$ of the $O_{H_{V,HFRep}}$ on the basis of already defined geometry:

\begin{enumerate}
    \item Depending on the nature of the attributes and how they are distributed in interior of the object $O_{H_{V,HFRep}}$, there are two possible types of object partitioning: single and multiple partitions. At this step we need to subdivide an object $O_{H_{V,HFRep}}$ according to the chosen partitioning scheme. 
    \newline
    \item Then we specify and evaluate an attribute function $F_{A_i}(F_{HFRep}(\bm{p}), \bm{p})$ for each partition to set up the attributes at the points $\bm{p} \in G$. These functions depend on the evaluation point coordinate and are parameterised by the computed distance using $F_{HFRep}(\bm{p})$ values.
    \newline
    \item In case when we have a multiple partitioned object with several specified attributes, we can obtain a single attribute function for all subsets $A_i$ by applying some interpolation, e.g. transfinite interpolation \cite{RVACHEV:2001} or space-time transfinite interpolation \cite{Sanchez:2015}.
\end{enumerate}
The more detailed discussion how to deal with attributes will be provided in section \ref{sec:basic_alg_attribs}.

\section{Algorithmic solutions for HFRep geometric shape generation}
\label{sec:detailed_basic_alg}

In this section we provide a detailed description of several particular steps of the basic algorithm outlined in the previous section. We consider a variety of combinations of FRep and SDF, ADF or IDF representations and propose a number of original solutions for solving problematic issues. The first step of the algorithm has been already discussed. In the next sections we discuss steps 2 - 4 (see Figs. \ref{fig:field_sequence}, (b) and (c)).

\subsection{Step 2: Generation of the unsigned distance field}
\label{ssec:hfrep_df_gen}

In this subsection we describe the solutions for generating UDFs. We show how some existing techniques can be used in this context and also introduce a novel method for the ADF generation.

\textbf{SDF generation.} To compute an approximate UDF, the most widely used class of methods is the distance transform (DT) \cite{Jones:2006}. DTs are efficiently generated on regular grids. In this work we suggest using the vector DT in which the vector components are propagated across the uniform grid. It provides a sufficiently accurate distance approximation. We follow the typical vector DT algorithm described in \cite{LEYMARI:1992} for 2D case and \cite{DANIELSSON:1980} for 3D case.

A definitive way to obtain an accurate DF for the object is to numerically solve the eikonal equation or the level-set PDEs \cite{Gomez:2019}. The numerical solution of PDE is quite time-consuming unless it is a multi-threading implementation of the method. The accuracy of the field is also highly dependent on the method. One of the robust methods for solving the eikonal equation is the fast iterative method (FIM) \cite{Jeong:2008}. It numerically solves a nonlinear Hamilton-Jacobi PDE defined on a Cartesian grid with a scalar speed function:
\begin{align}
    \label{eq:eikonal_general}
    H(\mathbf{p},\nabla \phi)&=|\nabla \phi(\mathbf{p})|^2 - \frac{1}{f^2(\mathbf{p})}=0, \quad \forall \mathbf{p} \in \mathbf{X} \subset \mathbb{R}^n\\
    \phi(\mathbf{p})&=0, \quad \mathbf{p} \in \Gamma \subset \mathbb{R}^n \nonumber
\end{align}
where $\mathbf{X}$ is a domain in $\mathbb{R}^n$, $\Gamma$ is the boundary condition, $\phi(\mathbf{p})$ is a travel time of the distance from the source to the grid point $\mathbf{p}$, $f(\mathbf{p})$ is a positive speed function and $H(\mathbf{p},\nabla \phi)$ is the Hamiltonian. The computed numerical solution is an unsigned distance on a uniform grid.
\begin{figure}[!t]
    \centering 
    \includegraphics[width=0.95\linewidth]{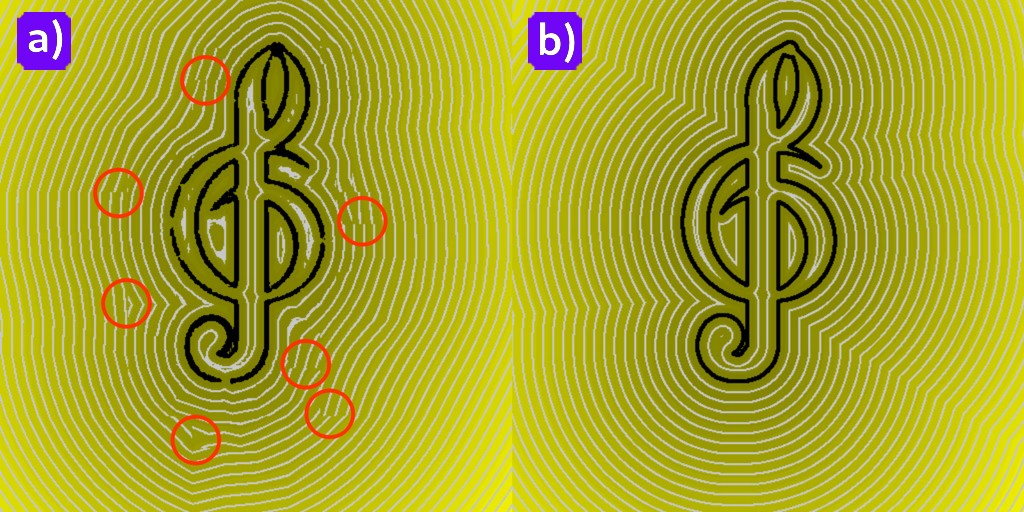}
    \caption{The comparison of the field restoration at each subdivided hierarchical cell using bilinear interpolation (a) and PHT-spline interpolation (b). In red circles we can see the $C^0$ discontinuity in the field isolines where the cells of different size appear next to each other.}
    \label{fig:bilinear_hbspline}
\end{figure}

\textbf{ADF generation.} To generate UDF on the hierarchical grid we briefly outline an original adaptation of the FIM method for solving the eikonal equation that also utilises PHT-splines \cite{Wang:2011} capability of the accurate geometry restoration. Our algorithm partly relies on the algorithm introduced in \cite{Jeong:2008} to inherit its advantages such as independent computation of each node and a simple data-structure (an active list $L$ or a doubly linked list) for handling node updates. A detailed description of the hierarchical FIM (HFIM) algorithm will be presented elsewhere.

The algorithm consists of two parts: (1) initialisation of the grid and (2) iterative updates of the numerical solution of the eikonal equation. First, we subdivide the space using quadtree/octree according to the values of the FRep field. We need to subdivide the exterior and interior of the FRep object with a small tree depth and its boundary with the maximum tree depth. While executing the hierarchical subdivision of the space, we also need to compute the basis functions for the PHT-splines \cite{Wang:2011} and reconstruct the PHT-spline surface that will be used for restoring distances in interior of each cell node.  

The idea of the hierarchical grid initialisation before applying HFIM is similar to the procedure described for FIM on the regular grid. We need to traverse the tree and set to zero those vertices of the cells that store the FRep values approximately equal to zero. The rest of the vertices are set to a relatively huge value. Thereafter, vertices that are equal to zero and the corresponding nodes are stored in the active list. The iterative computation of the solution of the eikonal equation on the hierarchical grid follows the logic of the FIM algorithm \cite{Jeong:2008}, but all steps are executed taking into account a hierarchical nature of the grid. The eikonal equation is iteratively solved using the first order upwind Godunov discretisation scheme that is modified for computations on the grid with irregular steps. The computed solution is stored and updated in all nodes that share the same vertices. The iterative computation is finished when the active list is empty. After obtaining the solution of the eikonal equation at each corner vertex of each cell of the hierarchical grid, we can restore the distance field using the already constructed PHT-spline surface.

As we have stated in subsection \ref{ssec:back_adfs}, the ADF field has $C^0$ discontinuities that arise after the hierarchical subdivision where cells of different size appear. In Fig. \ref{fig:bilinear_hbspline}, (a),   the discontinuities in the white isolines are located in the red circles. $C^1$ discontinuities are introduced by the bilinear/trilinear interpolation that is used for the field restoration in interior of each cell (see Fig. \ref{fig:bilinear_hbspline}, a). As it can be seen in  Fig. \ref{fig:bilinear_hbspline}, (b) the field generated by our method with PHT-spline restoration of the field successfully solves these drawbacks. All the isolines are continuous and smooth.

\begin{figure*}[!t]
    \centering
    \includegraphics[width=0.95\linewidth]{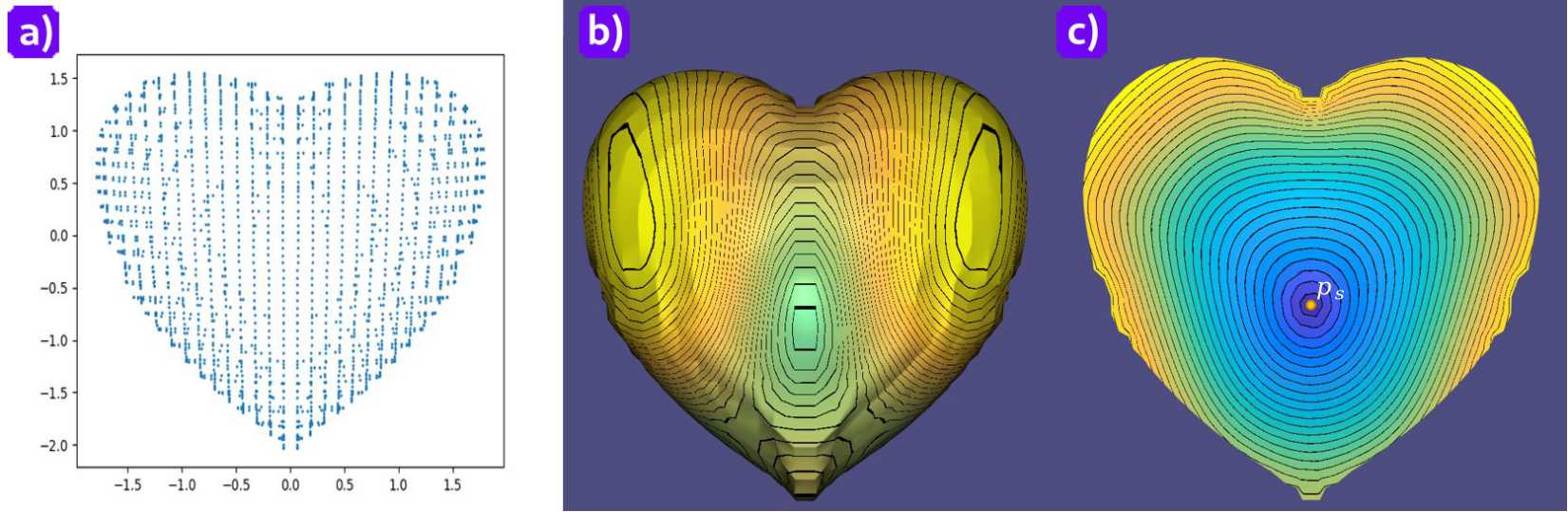}
    \caption {HFRep based on hybridisation of FRep and IDF generated for FRep 'heart' object using the method from \cite{Rustamov:2009}. a) the diffusion map computed on the surface of the object that is used for restoring distances at the shape boundary; b) the distances obtained at the boundary of the object shape that are shown as black isolines; c) is the tetrahedral slice of the mesh with isolines corresponding to the interior distances. The yellow point $\bm{p}_s$ corresponds to the 'source' point defined in the object interior.}
    \label{fig:hfrep_idf}
\end{figure*}

\textbf{IDF generation.} IDFs are usually computed using the solution of some PDE equations or, alternatively, some graph-based approach. We suggest using the approach described in \cite{Rustamov:2009}. 

The generation of IDFs is based on propagation of the distances computed on the boundary of the mesh in its interior. We will use Fig. \ref{fig:hfrep_idf} with the generated IDF field to explain how this method works. We start from triangulating an input geometric shape $G$ of the FRep object $O_{FRep}$ to generate the boundary surface $\partial G$ for further computations. The method, described in \cite{Rustamov:2009} was applied to tetrahedralised meshes and consists of two parts. 

First, we embed surface vertices $\bm{p}_{b_i}$ in some m-dimensional $\mathbb{R}^m$ space using a map $\bm{p}_{b_i} \mapsto \bm{p}_{b_i}^* \in \mathbb{R}^m$. This map was suggested to compute using diffusion maps introduced in \cite{COIFMAN:2006}. It can be obtained by computing an eigendecomposition $\{ \lambda_k, \phi_k\}_{k=1}^n$ of a discrete Laplace-Beltrami operator of the mesh. In Fig. \ref{fig:hfrep_idf}(a) we show the diffusion map obtained for the 'heart' object. The diffusion distances are computed as a Euclidean distance using obtained eigenvalues and eigenvectors \cite{deGoes:2008}.

After the diffusion distances were computed on the surface of the mesh as it can be seen in Fig. \ref{fig:hfrep_idf} (b), they are extended to the interior of the mesh using barycentric interpolation. If point $\bm{q} \in G_{in}$, then the barycentric representation of it is $\bm{q} \mapsto \bm{q}^*=\sum_i\omega_i(\bm{q})\bm{\nu}_i$, where $\omega_i(\bm{\cdot})$ are barycentric coordinates (e.g. mean-value coordinates in 2D \cite{Hormann:2006} or in 3D \cite{Schaefer:2005}). Finally, the distance in interior of the mesh can be obtained using computed diffusion distances $F^2_{DF}(\bm{p}_i,\bm{p}_j)$ and barycentric interpolation.

In Fig. \ref{fig:hfrep_idf} (c) we show a slice of the 'heart' object. The IDF was was computed between fixed 'source' point and the rest mesh points. One can see that the interior field is continuous, smoothly changing and following the boundary of the object.

At this step we can apply an extrapolation operation (e.g., using a wavenumber based extrapolation \cite{Tam:2000}) to the obtained IDF field to propagate the distances to exterior of the object using the already computed boundary distances. This operation will allow us to make the IDF field signed at the last step of the basic algorithm.

\subsection{Step 3: Smoothing an obtained UDF} 
\label{ssec:interpolation}

The resulting distance function $F_{DF}(\mathbf{p})$ for SDF, ADF or IDF is unsigned and satisfies the equation (\ref{eq:basic_df}). Having obtained UDF, we need to smooth the generated discrete field. To enforce an at least $C^1$ continuity and essential smoothness for the obtained field, we need to use some at least $C^1$ continuous interpolation function, e.g.  B-splines or bicubic/tricubic splines \cite{Knott:2000}.

\begin{figure}[!t]
    \centering
    \includegraphics[width=0.95\linewidth]{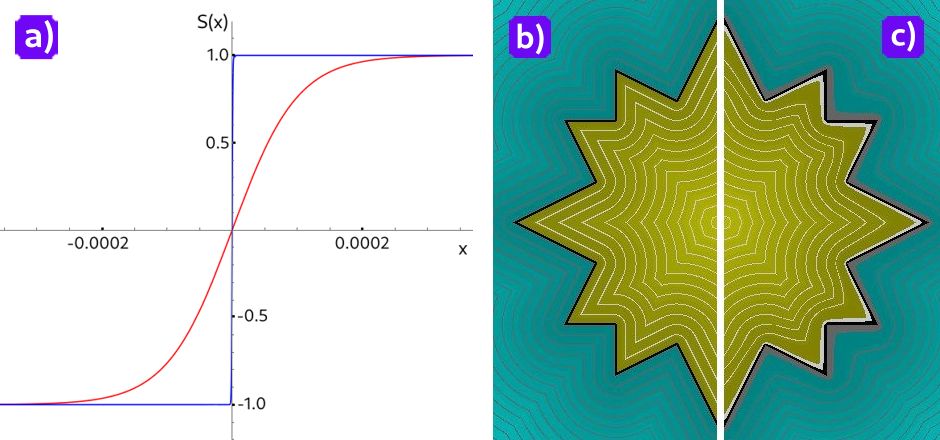}
    \caption{The illustration of the HFRep function continuity through throught varying the slope controlling parameter $s_l$. a) Plots of hyperbolic tangent sigmoid functions with according slope values $s_l$. Red line: $S_{sig}(x),\textit{ } s_l=10^{-4}$. Blue line: $S_{sig}(x), s_l=10^{-5}$; b) the HFRep 'star' object that was computed with $s_l = 0.00001$ for $F_{sig}$, equation (\ref{eq:step_fun}); c) the HFRep 'star' object that was computed with $s_l = 0.1$ for the $F_{sig}$, equation (\ref{eq:step_fun}); all sharp features are smooth, i.e. the HFRep function is $C^1$ continuous.}
    \label{fig:step_func}
\end{figure}

\subsection{Step 4: Distinguishing between interior, boundary and exterior of the object}
\label{ssec:sign}

At the previous step we had obtained a smooth and continuous unsigned distance function $F_{smDF}(\bm{p})$ defined by equation (\ref{eq:basic_inter}) that we used to compute UDF. Now, at the fourth step of the basic algorithm, we need to define the sign of UDF. To restore the sign we suggest to use a smooth step-function that depends on the values of the FRep function $F_{FRep}(\bm{p})$, defined at the first step of the basic algorithm. The step-function $F_{st}(F_{FRep}(\bm{p}))$ should satisfy the following requirements:

\begin{enumerate}
    \item it is approximately equal $-1$ when it corresponds to the exterior of the FRep object, $F_{FRep}(\bm{p})<0$; 
    \item it should be approximately equal to $0$ on the boundary of the FRep object, $F_{FRep}(\bm{p})=0$;
    \item it should be approximately equal to $1$ inside the FRep object, $F_{FRep}(\bm{p})>0$;
    \item it should be at least $C^1$ continuous everywhere in a Euclidean space $\mathbb{R}^n$;
    \item it should barely modify the values of UDF.
\end{enumerate}
\begin{figure}[!t]
    \center 
    \includegraphics[width=0.8\linewidth]{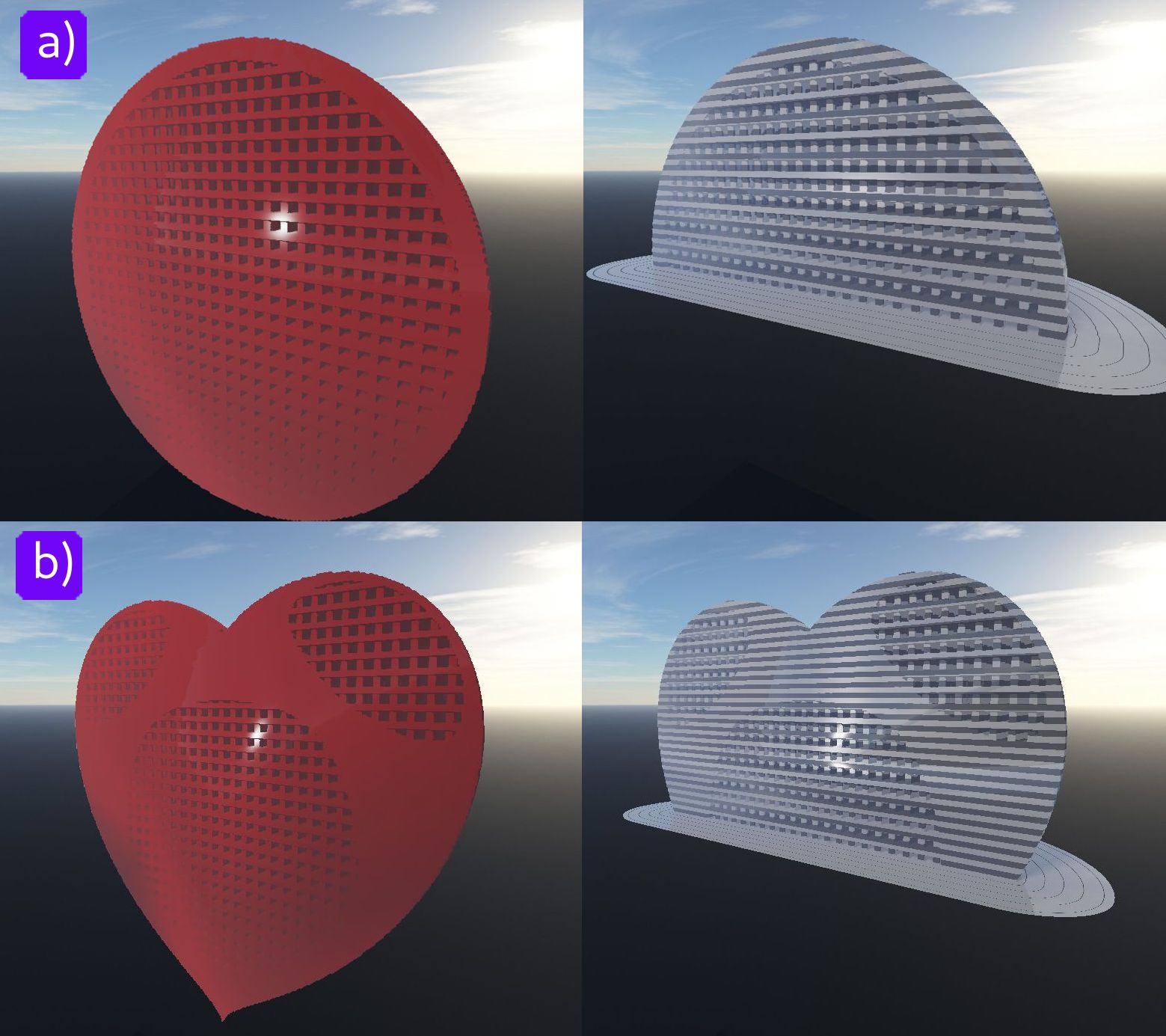}
    \caption{The illustration of the HFRep heterogeneous object based on the FRep and SDF representations with incorporated microstructure. a) the rendered HFRep 'sphere' object using the sphere-tracing method (left) and its isolines (right); b) the rendered HFRep 'heart' object using sphere-tracing method (left) and its isolines (right); }
   \label{fig:hfrep_microstr}
\end{figure}

We have identified two classes of functions which satisfy these requirements. These are sigmoid functions and spline functions, particularly cubic splines with Hermite end conditions, to estimate the slopes \cite{Knott:2000}. In this work we use the hyperbolic tangent sigmoid function \cite{Vogl:1988} (see Fig. \ref{fig:step_func}, a). By controlling slope parameter $s_l$, it is possible to get nearly step-function behaviour around zero:
\begin{align}
    \label{eq:step_fun}
    F_{sig}(x) &= \frac{r}{1+exp(-2x/s_l)} - \frac{r}{2}, \quad \forall x \in \mathbb{R}
\end{align}
where $r$ controls the range of the $F_{st}(x)$ along $y$-axes. We need to set parameter $r=2$ to make the function (\ref{eq:step_fun}) be defined in the interval $(-1, 1)$ along the y-axes.

The continuity of the HFRep function can be visualised as it is shown in Fig. \ref{fig:step_func} (b) and (c). In Fig. \ref{fig:step_func} (b), we show half of the 'star' object that was generated with $s_l=10^{-5}$ to follow the step-function shape as close as possible. In Fig. \ref{fig:step_func} (c), we show half of the 'star' object that was generated with $s_l=0.1$ to smooth the isolines shape. We can see that the $C^1$ continuity of the generated distance field is preserved and the obtained geometric shape of the object is watertight. 

\section{Dealing with attributes in HFRep framework}
\label{sec:basic_alg_attribs}

\begin{figure}
    \center 
    \includegraphics[width=0.95\linewidth]{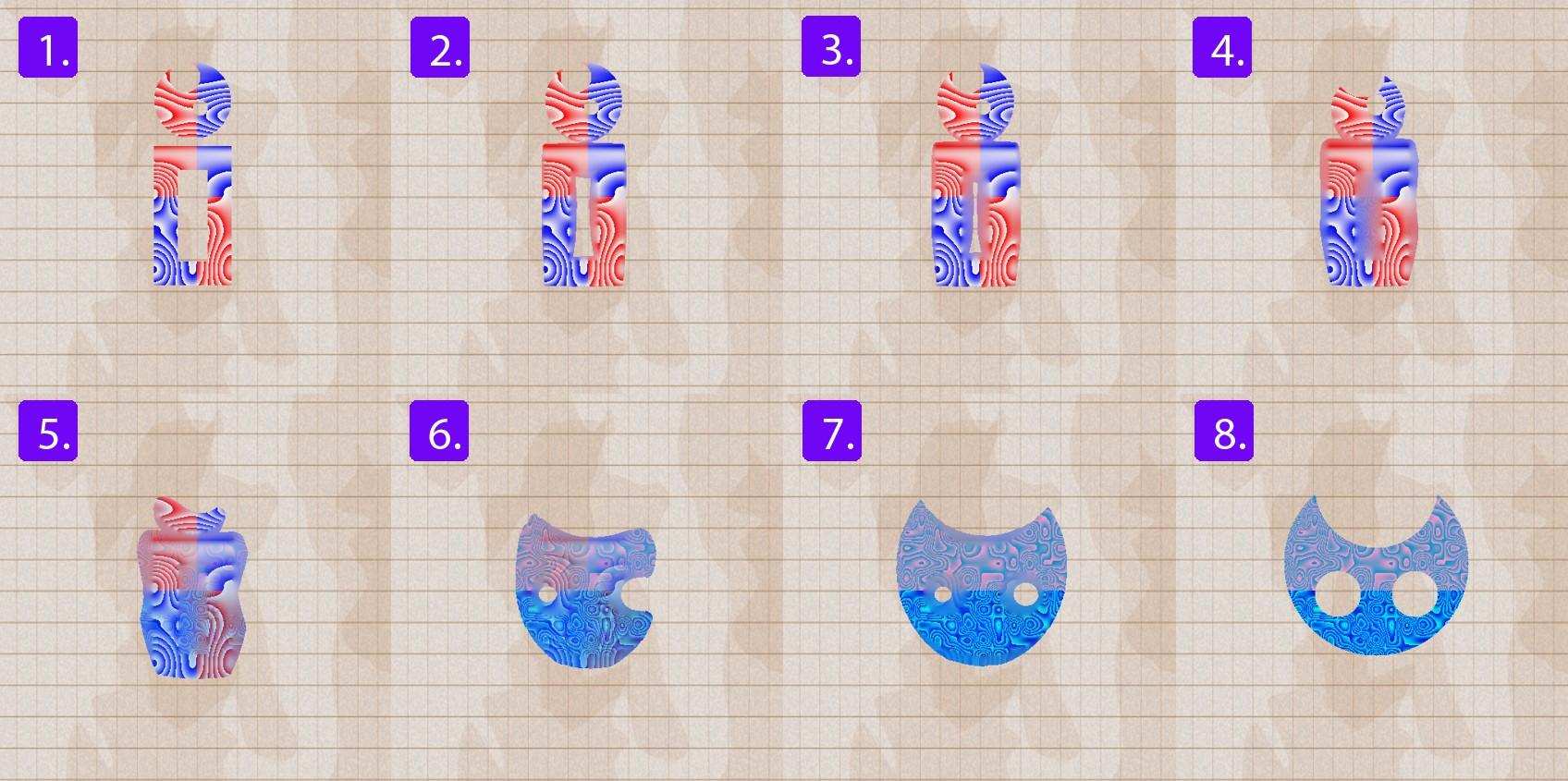}
    \caption{The illustration of the metamorphosis between two HFRep textured objects using the STB and STTI techniques. The texturing was made using procedural noise functions.}
   \label{fig:proc_texturing}
\end{figure}

In this section we show how we practically work with HFRep heterogeneous objects in terms of their attributes. In section \ref{sec:hfrep_basic_alg} we have outlined the basic algorithm for generating HFRep attribute functions. However, there is no universal approach for dealing with  HFRep object attributes because of their widely various nature. In this section we show how the proposed framework works for some representative attributes, namely, microstructures, colour and material attributes. We show the microstructures (Fig. \ref{fig:hfrep_microstr}), a heterogeneous model of the COVID-19 virus cell (Fig. \ref{fig:covid19}) and two models of metamorphosis dealing with a dynamic (time-variant) smooth transition from one HFRep object to another (Figs. \ref{fig:proc_texturing} and \ref{fig:single_attr_f}).

\subsection{Microstructures}
\label{ssec:microstr_ex}

In Fig. \ref{fig:hfrep_microstr} we demonstrate how microstructures in interior of the $O_{HFRep}$ object are implemented. The microstructures were defined as incorporated infinite slabs in interior of the 'sphere' and 'heart' objects using set-theoretic operations (\ref{eq:R0_sys}). The infinite slabs were defined according to \cite{PASKO:2011} as follows:
\begin{align}
    \label{eq:microstr_slabs}
    S(\bm{p})=\sin (\bm{\nu} \odot \bm{p} + \bm{\phi}) + \bm{l};
\end{align}
where $S(\bm{p})\geq 0$ is a vector function, with components defined as a set of slabs orthogonal to either X or Y or Z-axes, $\bm{\nu}$ is a frequency vector, with components defined as the distance between parallel slabs along one of the axes, $\bm{p}$ is a point $\bm{p} \in X$, $\bm{\phi}$ is a phase vector, with components defined as the position of slabs on one of the axes with respect to the origin and $\bm{l}$, $-1 < l_i < 1$ is a threshold vector that together with frequency parameters controls the thickness of each slab. Then the basic algorithm was applied to the obtained function to compute the HFRep objects with microstructures. 

Implementation was done using C++ and OpenGL. The HFRep geometric shape was computed as a scalar field which was stored in a 3D texture. Then it was passed into a fragment shader for assigning a single colour attribute and rendered using the sphere-tracing method.

\subsection{Procedural textures}
\label{ssec:proc_textures}

We can specify attributes as simple procedural textures. Fig. \ref{fig:proc_texturing} shows two heterogeneous HFRep objects $O_{H_v, HFRep}$ with coloured wooden textures that were obtained using a procedural function $f_{wood}(\bm{p})$. This function is constructed using hash table $htab(\bm{p})$ allowing for random sampling of the position values $\bm{p}$ multiplied by the frequency $\nu$. The procedural function for the wood can be defined as follows:
\begin{align}
\label{eq:proc_wood}
    g(\bm{p}) &= htab(\bm{p}\cdot \nu)\cdot c;
    \\
    f_{wood}(\bm{p}) &= g(\bm{p}) - int(g(\bm{p})); \nonumber
\end{align}
where $c>1$ is a constant, $g(\bm{p})$ is a noise function, $int(g(\bm{p}))$ is an integer part of the function $g(\bm{p})$ output value. To parameterise $f_{wood}(\bm{p})$ by the distance, we assign the distance values to the frequency parameter $\nu$.

Then a simple segmentation of the geometric shape of the objects was done (see Fig. \ref{fig:proc_texturing},(1)). We split the shape into four regions and assign colours using the obtained HFRep distance function $F_{HFRep}(\bm{p})$ and procedural function $f_{wood}(\bm{p})$ that defines the texture of the wood. The generated objects were used as inputs for 2D heterogeneous metamorphosis on the basis of the space-time blending (STB) method to handle geometry transformation and the space-time transfinite interpolation (STTI) to handle colour transformation \cite{Tereshin:2020}. This example was implemented using C++ and OpenCV.

\begin{figure}
    \center 
    \includegraphics[width=0.95\linewidth]{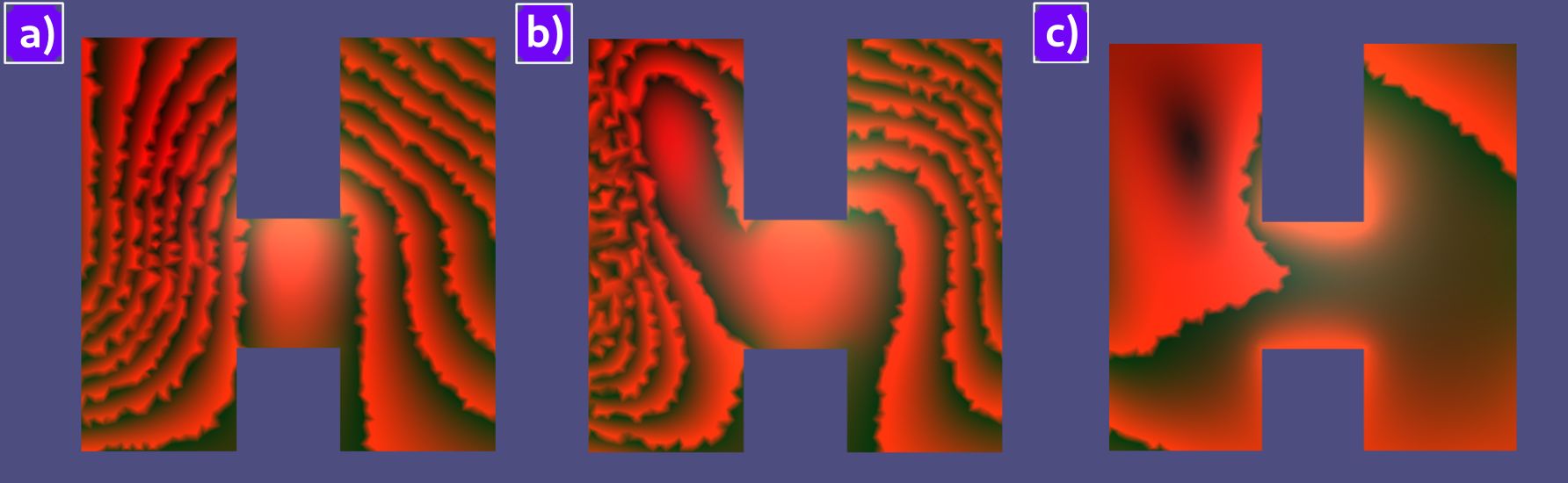}
    \caption{The HFRep 'H' object that is textured using procedural function \ref{eq:proc_wood} modelling the 'wood' texture. This function was differently parameterised (a), (b), (c) by computed IDF for the given object.}
   \label{fig:idf_param_tex}
\end{figure}

In another Fig. \ref{fig:idf_param_tex} we show three textured 'H' HFRep objects. The textures for these objects were generated using three different parameterisations of the procedural function for the wood by the computed IDFs.

\subsection{Voxel based attributes}
\label{ssec:voxel_based}

In this subsection we discuss how HFRep objects with voxel-based attributes can be defined. Two following examples were implemented in SideFX Houdini using the OpenVDB library.

\begin{figure}
    \center 
    \includegraphics[width=0.95\linewidth]{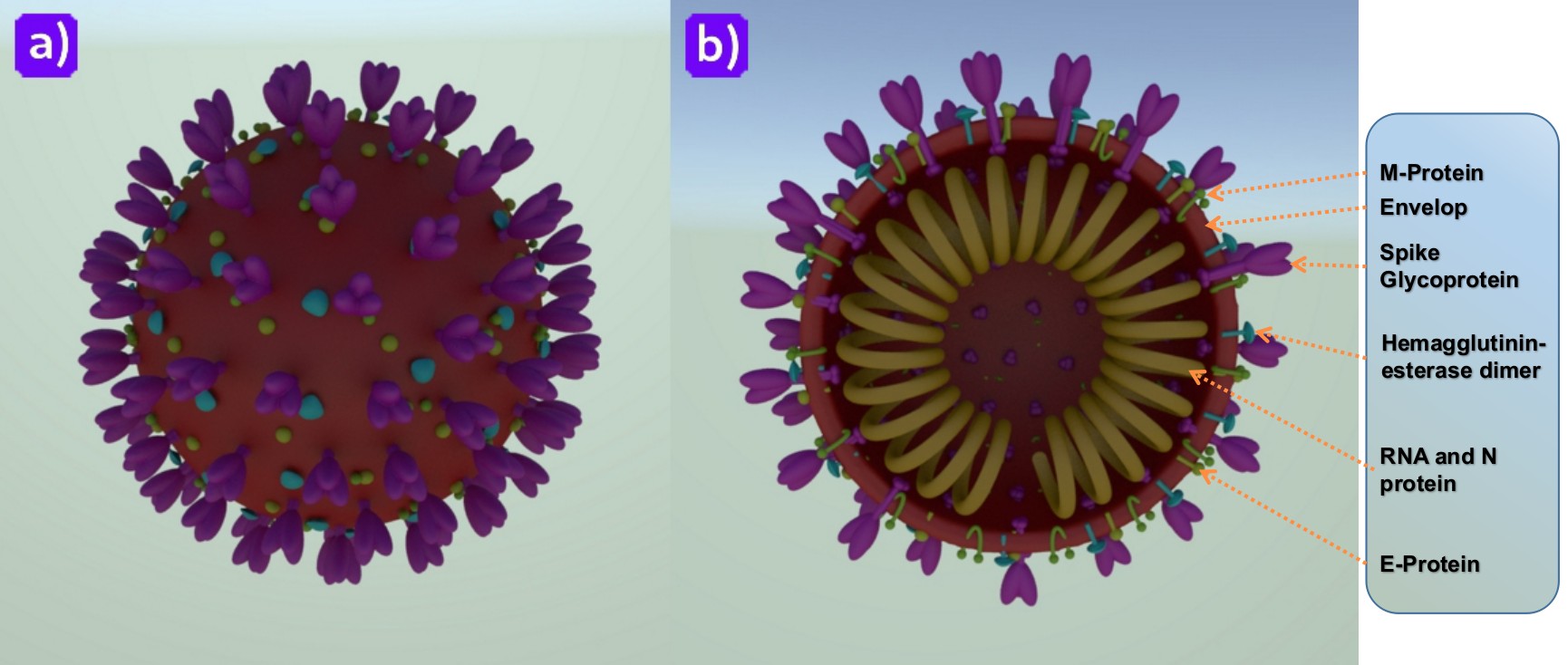}
    \caption{The COVID-19 cell model obtained as a combination of the HFRep and SDF functions. a) exterior of the virus cell; b) interior of the virus cell.}
   \label{fig:covid19}
\end{figure}

In Fig. \ref{fig:covid19} we show a 3D model of the COVID-19 cell that was obtained using 207 set-theoretic operations. In Fig. \ref{fig:covid19}, (b), we can see the interior structure of the COVID-19 cell \cite{MOUSAVIZADEH:2020}. The central part representing the RNA and N-protein was defined using SDF that was further combined with the HFRep spherical shell of the cell. The M-protein was also defined as a combination of the SDF arc and two HFRep spheres. The rest of the elements were defined using HFRep. Each element is mono-coloured and colours are assigned per-voxel.

\begin{figure*}[!t]
    \center 
    \includegraphics[width=1.0\linewidth]{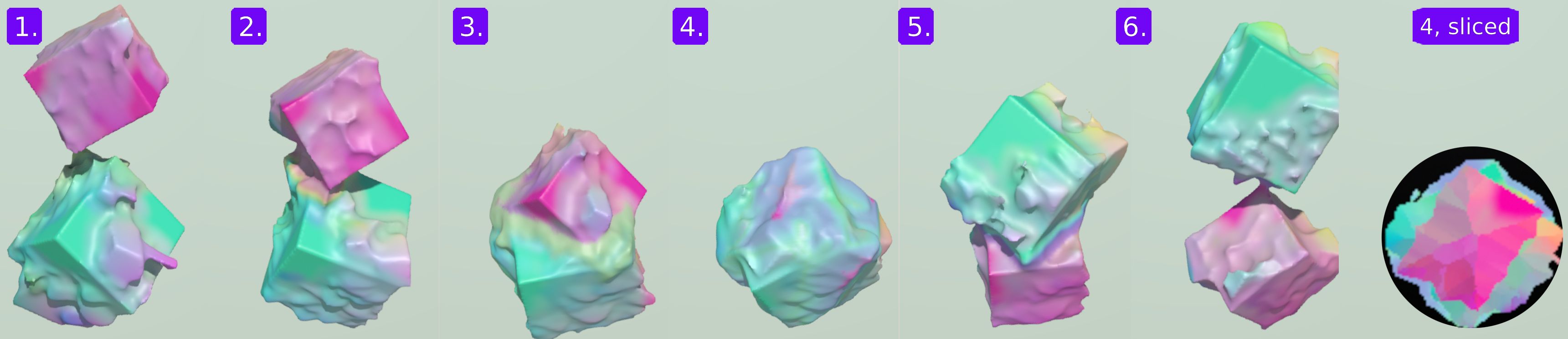}
    \caption{Metamorphosis between multiple coloured objects using the STB and STTI techniques. Colours of the initial objects are procedurally defined per voxel.}
   \label{fig:single_attr_f}
\end{figure*}

Fig. \ref{fig:single_attr_f} demonstrates a 3D metamorphosis between two heterogeneous objects that are a combination of the HFRep and SDF defined objects \cite{Tereshin_EG:2020}. This example served as one of tests for the '4D Cubism' project \cite{Corker-Marin:2018}. The input and target objects are two SDF cubes spaced from each other. These input shapes were segmented using an octree data-structure to make it possible a local faceting and distortions. Two colours were assigned to them per-voxel. Then different HFRep and SDF 'cubist' features were assigned to selected areas of the two basic SDF cubes, which were coloured per-voxel as randomly chosen colours from the specified range. Then we apply the same combination of methods as we have discussed before for the 2D metamorphosis. The generated colour and geometric shape transformations happen simultaneously and interconnectedly. In Fig. \ref{fig:single_attr_f}, 4 (sliced), we show how the interior of the object is transformed during the 3D metamorphosis process.

\section{Conclusions}
In this work we have introduced a theoretical and practical framework for modelling volumetric heterogeneous objects on the basis of a novel unifying functionally-based hybrid representation called HFRep. First, we have identified  four conventional representational schemes related to scalar fields of different kinds, namely FRep, SDF, ADF and IDF, suggested a formalisation of those approaches and described their advantages and drawbacks. This has allowed us to formulate the requirements for a unifying hybrid representation. The defining functions in the core of HFRep are continuous and have a distance property everywhere in a Euclidean space. They also have several other useful properties. We have defined the mathematical basics of the representation and developed an algorithmic procedure allowing to generate HFRep objects in terms of their geometry and attributes. 

To make our approach practical, we have provided a detailed description of the main steps of the algorithm and identified some problematic issues associated with them. This has required employing a number of techniques of different nature, separately and in combination. Some of these techniques were already described in literature, others had to be improved or developed. In particular, a new FIM algorithm for solving the eikonal equation on hierarchical grids has been developed. 

To show how the proposed framework works, we have illustrated the algorithmic process with a number of implemented examples, including those that deal with colour, material and microstructure attributes in the interior of functionally-defined shapes in the context of time-variant modelling.

While the boundary representation will remain the main and prevailing instrument for geometric  modelling, we believe that the functionally-based representations generalising a well-established implicit modelling approach, are becoming more important in the context of some modern applications. Hopefully, HFRep that embraces advantages and circumvents drawbacks of FRep, SDF, ADF, IDF will find its applications. 

Future work will be concerned with developing  operations over HFRep objects in the context of different applications, especially related to physical simulation, additive manufacturing and visual effects. In technical terms, we aim to develop a more efficient HFRep field extrapolation procedure beyond the computing domain. One of the interesting directions will be the introduction of the attribute definition in the interior of the volumetric object using the diffusion-based IDFs. We also consider a further generalisation of the FIM method for 3D hierarchical grids.

\bibliographystyle{unsrt}  
%\bibliography{references}  

\end{document}